\documentstyle[preprint,aps,epsf]{revtex}

\begin{document}
\tighten
\newcommand{\bea}{\begin{eqnarray}}
\newcommand{\eea}{\end{eqnarray}}
\draft

%\preprint{winnipeg }

\title {Shear viscosity in $\phi^4$ theory from an extended ladder resummation}

\author{ M.E. Carrington${}^{a,b}$, Hou Defu${}^{a,b,c}$ and R. Kobes${}^{b,d}$}

 \address{ ${}^a$ Department of Physics, Brandon University, Brandon, Manitoba,
R7A 6A9 Canada\\
 ${}^b$  Winnipeg Institute for Theoretical Physics, Winnipeg, Manitoba \\
${}^c$ Institute of Particle Physics, Huazhong Normal University, 430070 Wuhan,
China \\
${}^d$ University of Winnipeg, Winnipeg, Manitoba, R3B 2E9 Canada }

%\author{}
%\address{}

\date{\today}
\maketitle

\begin{abstract}

We study shear viscosity in weakly coupled hot $\phi^4$ theory using the closed
time path formalism (CTP) of real time finite temperature field theory. We show
that the viscosity can be obtained as the integral of a three-point
function. Although the three-point function has seven components in the CTP
formalism, we show that the viscosity is given by a
decoupled
integral equation which involves only one retarded three-point function.
Non-perturbative corrections to the bare one loop result can be obtained by
solving a Schwinger-Dyson type integral equation for this vertex. This
integral equation represents the resummation of an infinite series of ladder
diagrams which all contribute to the leading order result. It can be shown that
this integral equation has exactly the same form as the Boltzmann equation. We
show that the integral equation for the viscosity can be reexpressed by writing
the vertex as a combination of polarization tensors. An expression for this
polarization tensor can be obtained by solving another Schwinger-Dyson type
integral equation. This procedure results in an expression for the viscosity
which represents a non-perturbative resummation of contributions to the
viscosity which includes certain non-ladder graphs, as well as the usual
ladders. We discuss the significance of this set of graphs. We show that
these resummations can also be obtained by writing the viscosity as an integral
equation involving a single four-point function. Finally, we show that when
the viscosity is expressed in terms of a four-point function, it is possible to
further extend the set of graphs included in the resummation by treating vertex
and propagator corrections self-consistently. We discuss the significance of
such a self-consistent resummation and show that the integral equations that
are involved contain cancellations between vertex and propagator corrections.
Lastly, we discuss the prospect of generalizing our technique to gauge
theories.

\end{abstract}


\narrowtext
%%%%%%%%%%%%%%%%%%%%%%%%%%%%%%%%%%%%%%%%%%%%%%%%%%%%%%%%%%%%%%%%%%

\section{Introduction}

\label{sec1}

%%%%%%%%%%%%%%%%%%%%%%%%%%%%%%%%%%%%%%%%%%%%%%%%%%%%%%%%%%%%%%%%%%

%%%%%%%%%%%%%%%%%%%%%%%%%%%%%

The investigation of the transport properties of hot dense matter is a topic of
great interest in the context of relativistic  heavy-ion collisions
and astrophysics\cite{eh,blz}. The application of kinetic theories to the systems which are
relativistic and quantum in nature involves difficulties which have been only
partially overcome\cite{Groot}. There are two methods that are used to calculate
transport coefficients. The first method involves solving the transport
equations.  Transport coefficients
are calculated from classical transport equations using the relaxation time
approximation\cite{gyu}. The second method is to use the Kubo formulae which relate
transport coefficients to the  low frequency, zero momentum limits of the
spectral
densities of the appropriate composite operators\cite{kubo,MartinK,Hosoya,Horsley}. The Kubo formulae provide a
framework
for calculating transport coefficients starting from first
principles using finite temperature quantum field theory.

The diagrammatic rules
for computing  spectral densities have been derived in both the
real time formalism \cite{rdcut} and the imaginary time formalism\cite{jeon1}
of finite temperature field theory.
 One-loop calculations of transport coefficients using Kubo
formulae with effective propagators (including the single
particle thermal lifetimes) have appeared previously\cite{jeon1,heinz}.
However, these calculations are incomplete, even in the weak
coupling limit, because they do not include an infinite class
of planar ladder diagrams, which contribute at the same order as
the one-loop diagrams. The resummation of this infinite series of
ladder diagrams is highly nontrivial and has been performed
by Jeon \cite{jeon1,jeon2} using the imaginary time formalism.
In the imaginary formalism, this resummation is extremely complicated. In this
paper we perform the ladder resummation  in a more efficient way using the
closed time path (CTP) framework\cite{keld,schw,Chou,rep-145,Hu,PeterH}. We give a compact derivation of the
resummation of the infinite series of ladder diagrams that contribute to the
shear
viscosity. We can write the viscosity as an integral equation involving a
single three-point function which can be obtained as the solution of a further
decoupled linear integral equation. We show that these results are the same as
those obtained in \cite{jeon1}. We will also show that the ladder resummation
can be obtained by writing the viscosity as an integral equation involving a
single four-point function, which is obtained, as before, from a decoupled
linear integral equation. The four-point function is introduced because it
allows us to obtain a special type of self-consistent resummation which
includes additional non-ladder graphs. It is important to note that in the CTP
formalism the three-point function has seven components and the four-point function has fifteen
components, which means that the fact that these
integral equations are decoupled is highly non-trivial.

We also study extended resummations which include non-ladder contributions. We
derive expressions for the three- and four-point functions in terms of a
difference of self energies. By writing the self energy as the solution of a
decoupled linear integral equation we obtain an extended resummation. We
discuss the importance of non-ladder contributions to this resummation.

Finally, we show that when the viscosity is expressed in terms of a four-point
function, it is possible to perform a self-consistent resummation in which
corrections to the vertex and the propagators are put on the same footing. We
find there are cancellations between vertex and propagator corrections. For
$\phi^4$ theory, these cancellations do not significantly simplify the form of
the integral equation, however, for a gauge theory with a three-point
interaction, like scalar QED, we expect that the resulting integral equation
will be considerably simplified. We present the motivation for this argument. In
the last section we discuss our conclusions.

This paper is organized as follows.
\begin{itemize}
\item[] {\bf I}. Introduction
\item[] {\bf II}. Notation
\begin{itemize}
\item[] {\bf II-A}. The CTP Formalism
\item [] {\bf II-B}. Definition of Viscosity
\end{itemize}
\item[] {\bf III}. Viscosity From Corrected Three-Point Functions
\begin{itemize}
\item[] {\bf III-A}. Notation
\item[] {\bf III-B}. Integral equation for viscosity
\end{itemize}
\item[] {\bf IV}. Viscosity From Corrected Four-Point Functions
\begin{itemize}
\item[] {\bf IV-A}. Notation
\item[] {\bf IV-B}. Integral equation for viscosity
\end{itemize}
\item[] {\bf V}. Ladder Resummations
\begin{itemize}
\item[] {\bf V-A}. Ladders from three-point vertices
\begin{itemize}
\item[] {\bf V-A-1} The infrared divergence
\item[] {\bf V-A-2} The Boltzmann Equation
\end{itemize}
\item[] {\bf V-B}. Ladders from four-point vertices
\end{itemize}
\item[] {\bf VI}. Another Way to Resum Graphs
\begin{itemize}
\item[] {\bf VI-A}. The corrected three-point function
\begin{itemize}
\item[] {\bf VI-A-1}. The splitting relation for the three-point function
\item[] {\bf VI-A-2}. An integral equation for $\Pi$
\item[] {\bf VI-A-3}. Contributions to the viscosity
\item[] {\bf VI-A-4}. The infrared divergence
\end{itemize}
\item[] {\bf VI-B}. The corrected four-point function
\begin{itemize}
\item[] {\bf VI-B-1}. The splitting relation for the four-point function
\item[] {\bf VI-B-2}. An integral equation for $\Pi$
\item[] {\bf VI-B-3}. Contributions to the viscosity
\item[] {\bf VI-B-4}. The infrared divergence
\end{itemize}
\end{itemize}
\item[] {\bf VII}. Self Consistent Resummations
\item[] {\bf VIII}. Conclusions
\item[] Appendix A
\item[] References
\item [] Figures
\end{itemize}

%%%%%%%%%%%%%%%%%%%%%%%%%%%%%%%%%%%%%

\section{Notation}

%%%%%%%%%%%%%%%%%%%%%%%%%%%%%%%%%%%%%%%%%%%%%%%%%%%%%%%%%%%%%%%%%%

 \subsection{The CTP formalism}

The CTP formalism of finite temperature field theory was introduced by Keldysh
\cite{keld} and Schwinger \cite{schw}. Good reviews are found in
\cite{rep-145,Chou,PeterH}. The CTP contour has two branches: ${\cal C}_1$ runs
from negative infinity to positive infinity just above the real axis, and
${\cal C}_2$ runs back from positive infinity to negative infinity just below
the real axis. All fields can take values on either branch of this contour,
which results in a doubling in the number of degrees of freedom. We will consider $\phi^4$ theory.  The Lagrangian is given by,
\bea
{\cal L} = \frac{1}{2}(\partial_\mu \phi)^2 - \frac{1}{2} m^2 \phi^2 - \frac{\lambda}{4!}\phi^4 \nonumber
\eea
At finite temperature, when hard thermal loop corrections are included, the mass term is usually dropped relative to the real part of the hard thermal loop self energy. 
The scalar propagator is given by,
\bea D(X-Y) = -i\langle T_c \phi(X) \phi(Y) \rangle
\eea
where $T_c$ is the operator that time orders along the CTP contour. We also use
the notation $X=(t,\vec{x})$ and $P=(p_0,\vec{p})$.
The propagator has $2^2=4$ components and can be written as a $2 \times 2$
matrix
 \bea
 \label{2x2}
   D &=& \left(  \matrix {D_{11} & D_{12} \cr
              D_{21} & D_{22} \cr} \right) \nonumber
\eea
with
\bea
 D_{11}(X-Y) &=& -i\langle T(\phi(X) \phi(Y))\rangle \, , \nonumber\\
 D_{12}(X-Y) &=& -i\langle \phi(Y) \phi(X) \rangle \, , \nonumber\\
 D_{21}(X-Y) &=& -i\langle \phi(X) \phi(Y)\rangle \, , \nonumber\\
 D_{22}(X-Y) &=& -i\langle\tilde{T}(\phi(X)\phi(Y))\rangle \, ,\label{D11}
\eea
where $T$ is the usual time ordering operator, and $\tilde{T}$ is the
anti-chronological time ordering operator. These four components
satisfy,
 \begin{equation}
D_{11} - D_{12} - D_{21} + D_{22} = 0 \nonumber
\end{equation}
as a consequence of the identity $\theta(x) + \theta(-x) =1$.

It is more useful to write the propagator in terms of the three functions
\begin{eqnarray} \label{3a}
D_R &=& D_{11} - D_{12} \, , \nonumber\\
 D_A &=& D_{11} - D_{21} \, , \nonumber\\
 D_F &=& D_{11} + D_{22} \, .
\end{eqnarray}
$D_R$ and $D_A$ are the usual retarded and advanced propagators,
satisfying
 \begin{equation}
   D_R(X-Y)-D_A(X-Y) = -i\langle [\phi(X),\phi(Y)] \rangle\, ,\nonumber
 \end{equation}
and $D_F$ is the symmetric combination
 \begin{equation}
   D_F(X-Y) = -i\langle \{\phi(X),\phi(Y)\} \rangle\, ,\nonumber
 \end{equation}
which satisfies the KMS condition. In momentum space
\bea
  D_{R,A}(P) &&= \frac{1}{(p_0\pm i\epsilon)^2 - {\vec p}^{\,2}-m^2},\nonumber
\\
D_F(P) &&= (1+2n(p_0))(D_R(P) - D_A(P)),
\label{KMS1}
\eea
where $n(p_0)$ is the thermal Bose-Einstein distribution,
 \bea
 \label{Bose}
   n(p_0) = {1 \over e^{\beta p_0} -1}\, ,\qquad
   n(-p_0) = - \bigl( 1 + n(p_0) \bigr)\, ,~~~~~~~N_p = 1+2n(p_0)
 \eea
The propagator can be rewritten as an outer
product of two component column vectors \cite{PeterH}:
 \begin{equation}
 \label{decompD1}
   2\,D = D_R {1\choose 1}{1\choose -1}
        + D_A {1\choose -1}{1\choose 1}
        + D_F {1\choose 1}{1\choose 1}.
\end{equation}
Using the KMS condition~(\ref{KMS1}) this expression can be rewritten,
 \begin{eqnarray}
   D(p) &=& D_R(p) {1\choose 1}{1+n(p_0)\choose n(p_0)}
          - D_A(p) {n(p_0)\choose 1+n(p_0)}{1\choose 1}\, .
 \label{Dp}
 \end{eqnarray}

We can extract the 1PI two-point function, or self energy, by removing external
legs in the usual way. We find,
\bea
\Pi_R &=& \Pi_{11} + \Pi_{12} \nonumber \\
\Pi_A &=& \Pi_{11} + \Pi_{21} \nonumber \\
\Pi_F &=& \Pi_{11} + \Pi_{22} \label{physPi}
\eea
where $\Pi_R$ and $\Pi_A$ are the usual retarded and advanced self energies.
The symmetric component $\Pi_F$ satisfies the KMS condition,
\bea
\Pi_F = (1+2n(p_0))(\Pi_R - \Pi_A) \nonumber
\eea
and the four CTP components satisfy the constraint,
\bea
\Pi_{11} + \Pi_{12} + \Pi_{21} + \Pi_{22} =0 \nonumber
\eea

%Their spectral representations are

% \begin{equation}

% \label{spectral}

%   D_R(p) = \int_{-\infty}^\infty \frac{d\omega}{2\pi}

%          {\rho_-(\omega,{\bf p}) \over

%           p_0 - \omega + i\epsilon} \, , \qquad

%   D_A(p) = D^*_R(p).

% \end{equation}

%$\rho_-(p)$ is the (real) spectral density in terms of which all

%propagator components can be expressed via spectral integrals.

\subsection{Definition of Viscosity}

If a system is slightly perturbed away from equilibrium, fluctuations will
occur.
The response to these fluctuations will be characterized by transport
coefficients, for instance shear and bulk viscosities in a system with no
conserved
particle number. The shear viscosity characterizes the diffusive relaxation of
transverse momentum density fluctuations and is proportional to the two body
elastic scattering mean free path. The bulk viscosity characterizes the
departure from equilibrium during a uniform expansion and is proportional to
the mean
free path for particle number changing processes. Since the bulk viscosity
involves particle number changing processes, its
calculation is more difficult. However, at high temperature in a weakly
coupled system, the shear viscosity is much
larger than the bulk viscosity. For simplicity we will consider only the shear
viscosity in this paper.

 Using standard linear response theory one may express the shear viscosity in
terms of the stress tensor-stress tensor correlation function. One obtains the
Kubo formula\cite{kubo,jeon1},
\begin{equation}
\label{kubo}
\eta={\beta\over 20}\lim_{q_0 \to 0}\lim_{\vec{q} \to 0}
\sigma(Q)
\end{equation}
where,
\bea
\sigma(Q) = \int d X\, e^{i Q\cdot (X-Y)} \langle \pi_{lm}(X)
\pi^{lm}(Y)\rangle \label{sigmafirst}\eea
and
\bea
\pi_{lm}(X) = \nabla_l \phi \nabla_m \phi - \frac{1}{3} \delta_{lm}(\nabla
\phi)^2 \label{ppi}
\eea
in $\phi^4$ theory. We substitute~(\ref{ppi}) into~(\ref{sigmafirst}) and use
Wick's theorem and~(\ref{D11}). We obtain,
\bea
\sigma(Q) = 2\int dK\, I_{lm}(k,k) {\bf D}_{12}(K) {\bf D}_{21}(K+Q)
I_{lm}(k,k) \label{15}
\eea
This result is shown diagrammatically in Fig. [1]. The factor $I_{lm}$ is
associated with the joining of two propagators and has the form,
\bea
I_{lm}(k,k) = k_l k_m - \frac{1}{3}\delta_{lm} k^2
\eea
where we have dropped the $q$ dependence because we intend to take $q$ to zero
at the end of the calculation. In the factors $I_{lm}$ no difficulties will
arise from dropping  the $q$ dependence at the beginning of the calculation.
%
%If the lines in Fig. [1] are full propagators, given by exact solutions to the
%full hierarchy of coupled Schwinger-Dyson equations, then this expression is
%exact. In practice, of course, we do not know how to obtain a general solution
%to the Schwinger-Dyson equations. 
%
If we attempt to work perturbatively, the
lowest order contribution to (\ref{15}) is the one loop graph, as shown in Fig. [1]. For calculational purposes, it is easier to
work with the combination of one loop diagrams shown in Fig. [2].  The first
is $\sigma$ and the second we call $\sigma'$. The difference of these graphs
is related to $\sigma$ as follows. Using the KMS condition~(\ref{KMS1}) in the
form,
\bea
(1+n(p_0)) D_{12}(P) = n(p_0) D_{21}(P)\nonumber
\eea
it is straightforward to show that,
\begin{equation}
 n(q_0)\sigma=  (1+n(q_0))\sigma'
\end{equation}
which means that we can obtain $\sigma$ from the difference $\sigma -\sigma'$:
\bea
\label{sigma}
\sigma =  (1+ n(q_0))(\sigma - \sigma')\label{etap}
\eea

It has been known for some time that there are graphs that are higher order in
the loop expansion than the one loop graph which, nevertheless, contribute to
the same order in perturbation theory.
The purpose of this paper is to identify these graphs, and to develop
techniques for resumming them.

\section{Viscosity from corrected three-point functions}

\subsection{Notation}

We define the bare three-point vertex as,
\bea
\Gamma^{(0)lm}_{cba} = {\bf 1}_{c}\tau^3_b{\bf 1}_a \delta_{ab} \delta_{cb}
I^{lm}(k,k)
\label{GGamma0}
\eea
where ${\bf 1}$ is the two by two identity matrix and  $\tau_3$ is the third
Pauli matrix.
Corrections to this vertex are given by the 1PI part of the three-point
function,
\bea
\Gamma^{Clm}_{cba}(Z,X,Y) = \langle T_c \phi_c(Z)\tau^3_b \pi^{lm}(X) \phi_a(Y)
\rangle \label{GGamma}
\eea
In the real time formalism, the three-point function has $2^3 = 8$
components, since each of the fields can take values on either branch of the
contour. Only seven of these components are independent because of the
identity
 \begin{equation}
   \sum_{a=1}^2\sum_{b=1}^2\sum_{c=1}^2
   (-1)^{a+b+c-3} \Gamma^{Clm}_{abc} =0.
 \label{eq: circVer}
 \end{equation}
The corrected three-point vertex is obtained by truncating external legs to
obtain the 1PI part of the three-point function. These 1PI vertices obey the
constraint,
\bea
\sum_{a=1}^2\sum_{b=1}^2\sum_{c=1}^2 \Gamma^{lm}_{abc} = 0 \nonumber
\eea
We define the combinations,
 \begin{eqnarray}
   {\Gamma}^{lm}_{R} &=& {\Gamma}^{lm}_{111} + {\Gamma}^{lm}_{112}
                + {\Gamma}^{lm}_{211} + {\Gamma}^{lm}_{212}  \nonumber\\
  {\Gamma}^{lm}_{Ri} &=& {\Gamma}^{lm}_{111} + {\Gamma}^{lm}_{112}
                   + {\Gamma}^{lm}_{121} + {\Gamma}^{lm}_{122}  \nonumber\\
   {\Gamma}^{lm}_{Ro} &=& {\Gamma}^{lm}_{111} + {\Gamma}^{lm}_{121}
                   + {\Gamma}^{lm}_{211} + {\Gamma}^{lm}_{221}  \nonumber\\
   {\Gamma}^{lm}_{F} &=& {\Gamma}^{lm}_{111} + {\Gamma}^{lm}_{121}
                + {\Gamma}^{lm}_{212} + {\Gamma}^{lm}_{222}  \nonumber\\
   {\Gamma}^{lm}_{Fi} &=& {\Gamma}^{lm}_{111} + {\Gamma}^{lm}_{122}
                   + {\Gamma}^{lm}_{211} + {\Gamma}^{lm}_{222}  \nonumber\\
   {\Gamma}^{lm}_{Fo} &=& {\Gamma}^{lm}_{111} + {\Gamma}^{lm}_{112}
                   + {\Gamma}^{lm}_{221} + {\Gamma}^{lm}_{222}  \nonumber\\
   {\Gamma}^{lm}_{E} &=& {\Gamma}^{lm}_{111} + {\Gamma}^{lm}_{122}
                + {\Gamma}^{lm}_{212} + {\Gamma}^{lm}_{221} .
 \label{eq: physVer2}
 \end{eqnarray}
 The first three are the retarded three-point
functions; $\Gamma^{lm}_{Ri}$ is retarded with respect to the first leg,
$\Gamma^{lm}_{R}$ is retarded with respect to the second leg, and
$\Gamma^{lm}_{Ro}$ is
retarded with respect to the third leg. The other four vertices are related to
the retarded ones through the KMS conditions \cite{u&m,hou2},
 \begin{eqnarray}
   \Gamma^{lm}_F &=& N_1 (\Gamma^{*lm}_R - \Gamma^{lm}_{Ro})
              + N_3 (\Gamma^{*lm}_R - \Gamma^{lm}_{Ri}) \, ,
 \nonumber\\
   \Gamma^{lm}_{Fi} &=& N_2 (\Gamma^{*lm}_{Ri} - \Gamma^{lm}_{Ro})
                  + N_3 (\Gamma^{*lm}_{Ri} - \Gamma^{lm}_R) \, ,
 \nonumber\\
   \Gamma^{lm}_{Fo} &=& N_1 (\Gamma^{*lm}_{Ro} - \Gamma^{lm}_R)
                  + N_2 (\Gamma^{*lm}_{Ro} - \Gamma^{lm}_{Ri}) \, ,
 \nonumber\\
   \Gamma^{lm}_E &=& \Gamma^{*lm}_{Ri} + \Gamma^{*lm}_R + \Gamma^{*lm}_{Ro}
              + N_2 N_3 (\Gamma^{lm}_{Ri} + \Gamma^{*lm}_{Ri})
 \nonumber\\
            &+& N_1 N_3 (\Gamma^{lm}_R + \Gamma^{*lm}_R)
              + N_1 N_2 (\Gamma^{lm}_{Ro} + \Gamma^{*lm}_{Ro}) \, ,
 \label{KMSver}
 \end{eqnarray}
The vertex can be written as the outer product of three column vectors:
 \begin{eqnarray}
   4\,{\Gamma}^{lm} &=& {\Gamma}^{lm}_R {1 \choose 1} {1\choose -1} {1\choose
1}
               + {\Gamma}^{lm}_{Ri} {1 \choose -1}{1\choose 1} {1 \choose 1}
 \nonumber\\
           &+& {\Gamma}^{lm}_{Ro} {1 \choose 1} {1\choose 1} {1 \choose -1}
             + {\Gamma}^{lm}_F {1\choose -1} {1\choose 1}{1\choose -1}
 \nonumber\\
           &+& {\Gamma}^{lm}_{Fi} {1\choose 1}{1\choose -1}{1\choose -1}
             + {\Gamma}^{lm}_{Fo} {1\choose -1}{1\choose -1}{1\choose 1}
 \nonumber\\
           &+& {\Gamma}^{lm}_E {1\choose -1}{1\choose -1}{1\choose -1}.
 \label{decompver}
 \end{eqnarray}

\subsection{Integral equation for viscosity}

In order to perform a resummation we look at an integral equation of the form
represented in Fig. [3]. In these diagrams, the vertex with the solid box is a full vertex
whose form will be specified later. As we will see, it is the form of this
vertex that determines which diagrams are being resummed.
The four diagrams with corrected vertices in Fig. [3] are necessary in order to
obtain a result for $\sigma$ that is pure real. Using corrected vertices on
the right and left sides alternately is necessary to correctly take cuts of the
vertex itself. Notice that diagrams with corrected vertices include an extra
factor of $\tau_3$ which carries the index that corresponds to the momentum
$Q$. The reason is as follows. The places where lines join in the bare one
loop diagrams are not vertices and do not carry any factors, numerical or
otherwise (expect for the kinematic factor $I_{lm}$). In the diagrams with
corrected vertices, the corrected vertex is given by~(\ref{GGamma}). The
factor of $\tau_3$ which is included in the definition of the vertex must be
compensated for by an extra $\tau_3$ which carries the index that corresponds
to the momentum $Q$. This extra factor of $\tau_3$ ensures that the diagram
with the full vertex would reduce to the bare one loop diagram in the tree
limit, which means that the combinatorics are correct in the integral equation
that derives from Fig. [3]. The integral equation corresponding to the
diagrams in Fig. [3a] is of the form,
\bea
\lambda_{ab}^{(1)} = 2\int dK\, I_{lm}(k,k) D_{ab}(K+Q) D_{ba}(K) I_{lm} (k,k)
\eea
Using the decomposition~(\ref{decompD1}) we obtain,
\bea
&&(\sigma-\sigma')^{(1)} = (\lambda_{21}-\lambda_{12})^{(1)}  \\
&&~~=  \int dK\,I_{lm}(k,k)[f_{k+q}(a_k - r_k) - f_k(a_{k+q} - r_{k+q})]I_{lm}
(k,k)
\nonumber
\eea
where we define $a_k=D_A(K), r_k=D_R(K), f_k=D_F(K)$ etc. Using the KMS
condition~(\ref{KMS1}) we obtain,
\bea
(\sigma-\sigma')^{(1)} = - \int dK\, I_{lm}(k,k) (N_k - N_{k+q}) (a_k r_{k+q} +
h.c.) I_{lm} (k,k) \label{1a}
\eea
In obtaining this result we have used the fact that we will eventually take the
limit $Q\rightarrow 0$ to obtain the viscosity~(\ref{kubo}). This limit gives
rise to what is known as the pinch effect:  terms with a product of factors
$a_k r_{k+q}$ and $r_k a_{k+q}$ contain an extra factor $\sim 1/Q$ relative to
terms with products of propagators $a_k a_{k+q}$ or $r_k r_{k+q}$. Thus terms
proportional to $a_k a_{k+q}$ or $r_k r_{k+q}$ can be dropped. (The large terms
occur when the contour is ``pinched'' between the poles of the two propagators,
which gives rise to a factor in the denominator that is proportional to the
imaginary part of these propagators).

The diagrams in Fig. [3b] correspond to an expression of the form,
\bea
2\lambda^{(2)}_{ab} &&=  \int dK\, \tau_3^a\Gamma^{lm}_{cad}(K,Q,-K-Q)
D_{db}(K+Q) D_{bc}(K) I^{lm}(k,k) \nonumber \\
&&~~~+  \int dK\, I_{lm}(k,k) D_{ac}(K+Q) D_{da}(K)
\Gamma^{lm}_{dbc}(-K,-Q,K+Q) \tau_3^b
\eea
Using the decompositions~(\ref{decompD1}) and~(\ref{decompver}) and the KMS
condition~(\ref{KMS1}) we obtain,
\bea
&&(\sigma-\sigma')^{(2)} = (\lambda_{21}-\lambda_{12})^{(2)} \\
&&~~= \frac{1}{4} \int dK\, \left\{ \right. a_k r_{k+q} [(N_{k+q}-N_k)
\Gamma^{lm}_R + N_{k+q}\Gamma^{lm}_{Ri} - N_k \Gamma^{lm}_{Ro} -
\Gamma^{lm}_F]I^{lm}(k,k)  \nonumber \\
&&~~~~~~ - r_k a_{k+q} I^{lm}(k,k) [ (N_k - N_{k+q}) \Gamma_R^{lm(-)} + N_k
\Gamma_{Ro} ^{lm(-)} - N_{k+q} \Gamma_{Ri}^{lm(-)} + \Gamma_F^{lm(-)}] \left.
\right\}\nonumber
\eea
where $\Gamma_x := \Gamma_x(K,Q,-K-Q)$ and $\Gamma_x^{(-)} :=
\Gamma_x(-K,-Q,K+Q)$ and  we have again used the pinch limit.
We can further simplify this result by using the KMS condition~(\ref{KMSver})
with the relations $\Gamma^{lm(-)}_{Rx} = [\Gamma_{Rx}^{lm}]^*$ and
$\Gamma^{lm(-)}_{Fx} = -[\Gamma_{Fx}^{lm}]^*$. We obtain,
\bea
(\sigma-\sigma')^{(2)} = \frac{1}{2} \int dK\,
\left\{\Gamma^{lm}_R(K,Q,-K-Q)a_k r_{k+q} + h.c.\right\}(N_{k+q} - N_k)
I_{lm}(k,k) \label{2b}
\eea
Combining~(\ref{1a}) and~(\ref{2b}) we obtain,
\bea
&&\sigma-\sigma' \label{int2} \\
&&=  \int dK\, \left\{ (I_{lm}(k,k) + \frac{1}{2}\Gamma^{lm}_R(K,Q,-K-Q))a_k
r_{k+q} + h.c.\right\} (N_{k+q} - N_k) I_{lm}(k,k) \nonumber
\eea
Notice that this expression depends only on the vertex $\Gamma^{lm}_R$.

\section{Viscosity from CORRECTED FOUR POINT FUNCTIONS}

It is also possible to obtain a ladder resummation by writing the viscosity as
an integral involving a corrected four-point function. This approach has
advantages over the three-point function  because it can be used to generate
self consistent resummations. This point will be explained in detail below.

\subsection{Notation}

The connected four-point function is given by the contour ordered expectation
value,
\bea
M^C_{abcd}(X,Y,Z,W) = \langle T_c \phi_a(X) \phi_b(Y) \phi_c(Z)
\phi_d(W)\rangle \nonumber
\eea
The 1PI four-point function is obtained by truncating external legs and forms a
16 component tensor which we can write as the
outer product of four two component vectors,
\bea
M = {x \choose y} {u\choose v} {w\choose z} {s\choose t} \nonumber
\eea

The retarded 1PI four-point functions are given by
\begin{eqnarray}
M_{R1} &=& M_{1111} + M_{1112} + M_{1121} + M_{1211} + M_{1122} + M_{1212} +
M_{1221} + M_{1222} \nonumber\\
   &=& \frac{1}{2} (x-y)(u+v)(w+z)(s+t)\nonumber \\
M_{R2} &=& M_{1111} + M_{1112} + M_{1121} + M_{2111} + M_{1122} + M_{2112} +
M_{2121} + M_{2122}\nonumber \\
   &=& \frac{1}{2} (x+y)(u-v)(w+z)(s+t) \label{Ms}\\
M_{R3} &=& M_{1111} + M_{1112} + M_{2111} + M_{1211} + M_{2112} + M_{1212} +
M_{2211} + M_{2212} \nonumber\\
   &=& \frac{1}{2} (x+y)(u+v)(w-z)(s+t)\nonumber\\
M_{R4} &=& M_{1111} + M_{2111} + M_{1121} + M_{1211} + M_{2121} + M_{2211} +
M_{1221} + M_{2221} \nonumber\\
   &=& \frac{1}{2} (x+y)(u+v)(w+z)(s-t)\nonumber
\end{eqnarray}
where we have used the relation
\begin{equation}
\sum_{a,b,c,d=1}^{2} M_{abcd} = 0.
\end{equation}
The other combinations are,
\begin{eqnarray}
M_A &=& \frac{1}{2}(x+y) (u+v) (w-z) (s-t)\nonumber \\
M_B &=& \frac{1}{2}(x-y) (u+v) (w-z) (s+t)\nonumber\\
M_C &=& \frac{1}{2}(x+y) (u-v) (w-z) (s+t) \nonumber\\
M_D &=& \frac{1}{2}(x+y) (u-v) (w+z) (s-t) \nonumber\\
M_E &=& \frac{1}{2}(x-y) (u-v) (w+z) (s+t) \nonumber\\
M_F &=& \frac{1}{2}(x-y) (u+v) (w+z) (s-t) \\
M_\alpha &=& \frac{1}{2}(x+y) (u-v) (w-z) (s-t)\nonumber\\
M_\beta &=& \frac{1}{2}(x-y) (u+v) (w-z) (s-t)\nonumber\\
M_\gamma &=& \frac{1}{2}(x-y) (u-v) (w+z) (s-t) \nonumber\\
M_\delta &=& \frac{1}{2}(x-y) (u-v) (w-z) (s+t) \nonumber\\
M_T &=& \frac{1}{2}(x-y) (u-v) (w-z) (s-t) \nonumber
\end{eqnarray}

We use the decomposition of the four-point vertex:
\bea
8M &&= M_{R1}{1\choose -1}{1\choose 1}{1\choose 1}{1\choose 1} + M_{R2}
{1\choose 1}{1\choose -1}{1\choose 1}{1\choose 1} + M_{R3} {1\choose
1}{1\choose 1}{1\choose -1}{1\choose 1} \nonumber \\
&&+ M_{R4} {1\choose 1}{1\choose 1}{1\choose 1}{1\choose -1} + M_A {1\choose
1}{1\choose 1}{1\choose -1}{1\choose -1} + M_B {1\choose -1}{1\choose
1}{1\choose -1}{1\choose 1} \nonumber\\
&&+ M_C {1\choose 1}{1\choose -1}{1\choose -1}{1\choose 1} + M_D {1\choose
1}{1\choose -1}{1\choose 1}{1\choose -1} + M_E {1\choose -1}{1\choose
-1}{1\choose 1}{1\choose 1}\label{decompM} \\
&&+ M_F {1\choose -1}{1\choose 1}{1\choose 1}{1\choose -1} + M_{\alpha}
{1\choose 1}{1\choose -1}{1\choose -1}{1\choose -1} + M_\beta {1\choose
-1}{1\choose 1}{1\choose -1}{1\choose -1} \nonumber\\
&&+ M_\gamma {1\choose -1}{1\choose -1}{1\choose 1}{1\choose -1} + M_\delta
{1\choose -1}{1\choose -1}{1\choose -1}{1\choose 1} + M_T {1\choose
-1}{1\choose -1}{1\choose -1}{1\choose -1}\nonumber
\eea

\subsection{Integral equation for viscosity}

The integral equation for the viscosity is obtained from Fig [4]. The one loop
diagrams give, as before~(\ref{1a}).
The second diagrams are of the form,  
\bea
\lambda_{ab}=\int dP\,dK\,&&I_{lm}(p,p) I_{lm}(k,k)\nonumber \\
&& D_{ac}(P+Q) D_{db}(K+Q) D_{bf}(K) D_{ed}(P) M_{cdef}(P+Q,-K-Q,-P,K)
\nonumber \\
\eea
We want to extract the combination $\lambda_{21}-\lambda_{12} = (\sigma -
\sigma')^{(2)}$. The result is,
\bea
(\sigma - \sigma')^{(2)} = \frac{1}{4}&&\int dP\,dK I_{lm}(p,p)
I_{lm}(k,k)\left[M_{R1}(r_{p+q} f_p (a_k f_{k+q} f_k r_{k+q})+ f_k f_{k+q}
r_{p+q}r_p )\right. \nonumber \\
&& -M_{R2}(f_k a_{k+q}(a_{p+q}f_p + f_{p+q}r_p))
+M_{R3}(f_{p+q}a_p(a_k f_{k+q} + f_k r_{k+q}))\nonumber \\
&& +M_{R4}(f_{p+q}r_k(f_pr_{k+q} - f_{k+q}r_p) - r_k f_{k+q} a_{p+q}
f_p)\nonumber \\
&& \left. +M_B(r_{p+q}a_p(a_kf_{k+q} + f_kr_{k+q})) - M_D(r_ka_{k+q}(r_pf_{p+q}
+ f_pa_{p+q}))\right] \label{intM1}
\eea
where we have suppressed the momentum arguments of the four-point functions.

To simplify this result, we take the pinch limit and use the KMS conditions.
These conditions for the four-point functions are analogous to the KMS
conditions for the three-point functions~(\ref{KMSver}) and are derived in
\cite{hou2}. The only one that we need is,
\bea
(N_k-N_{k+q})[M_B - N_p M_{R1} + N_{p+q} M_{R3}] = (N_p - N_{p+q})[M_D^* +
N_kM_{R2}^* - N_{k+q}M_{R4}^*]  \label{KMSM}
\eea
%\bea
%&&(N_k-N_{k+q})M_B + (N_p-N_{p+q})M_d^* \label{KMSM} \\
%&&= N_p(N_k-N_{k+q})M_{R1} - N_{p+q}(N_k-N_{k+q}) M_{R3} - N_k(N_{p+q}-N_p)
%%%M_{R2}^* + N_{k+q}(N_{p+q} - n_p) M_{R4}^* \nonumber
%\eea
Substituting~({\ref{KMSM}) into~(\ref{intM1}) we get,
\bea
(\sigma-\sigma')^{(2)}&& = -\frac{1}{4} \int dP\, dK\, I_{lm}(p,p) I_{lm}(k,k)
(N_k - N_{k+q} ) \nonumber \\
&&~~\left[ (M_B - N_p M_{R1} + N_{p+q} M_{R3}) r_{k+q} a_k r_{p+q} a_p +
h.c.\right]
\label{intM2}
\eea
Making the definition
\bea
&&\tilde{M}(P+Q,-K-Q,-P,K) = M_B(P+Q,-K-Q,-P,K) \nonumber \\
&&~~~~- N_p M_{R1}(P+Q,-K-Q,-P,K) + N_{p+q}M_{R3}(P+Q,-K-Q,-P,K) \label{Mtilde}
\eea
we can rewrite to obtain,
\bea
(\sigma-\sigma')^{(2)} = -\frac{1}{4} \int dP\, dK\, I_{lm}&&(p,p) I_{lm}(k,k)
(N_k - N_{k+q}) \left[ \tilde{M} r_{k+q} a_k r_{p+q} a_p + h.c.\right]
\label{intM2b}
\eea
Combining~(\ref{1a}) and~(\ref{intM2b}) we obtain from Fig. [4],
\bea
\sigma-\sigma' = - &&\int dK (N_k - N_{k+q}) I_{lm}(k,k) \left\{\right. a_k
r_{k+q} \nonumber \\
&&~~~~~~~\left[ I_{lm}(k,k) + \frac{1}{2} \int dP\,I_{lm}(p,p) \tilde{M} a_p
r_{p+q} \right] + h.c.\left.\right\}
\label{intM4b}
\eea

\section{Ladder resummations}

An argument similar to the pinch argument described in the section III-B 
gives rise to the conclusion that ladder graphs of the form shown in Fig. [5]
contribute to the viscosity to the same order in perturbation theory as the
bare one loop graph, and thus need to be resummed \cite{jeon1,jeon2}. This
effect occurs for the following reason. It appears that the ladder graphs are
suppressed relative to the one loop graph by extra powers of the coupling,
which come from the extra vertex factors that one obtains when one adds rungs
(vertical lines). However, these extra factors of the coupling are compensated
for by a kinematical factor $\sim 1/Q$ from the pinch effect, which becomes
large when $Q$ becomes small. One can also argue that the ladder graphs are
larger than other graphs of the same order in the loop expansion. This
conclusion is based on the following argument. We compare the graphs shown in
Fig. [6a] and [6b]. The ladder graph shown in Fig. [6a] contains three pairs
of propagators of the form $D_x D_{x+q}$ which will give pinch contributions in
the limit that $Q\rightarrow 0$. The graph in Fig. [6b] contains only two such
pairs; the third pair is replaced by a pair of the form $D_{x+y} D_{x+q}$. The
momentum variable $y$ is an internal momentum which must be integrated over and
thus there is a region of phase space for which it will be small. However, one
expects that the measure will also be small in this region of phase space, and
therefore that the contribution of all non-ladder graphs will be suppressed.
This argument will be discussed in more detail in the next section. In this
section, we will show that we can resum the ladder graphs by solving an
integral equation for the corrected vertex.

\subsection{Ladders from three-point vertices}

The ladder graphs are resummed by solving an integral equation for the
three-point vertex.  We can obtain
 $\Gamma^{lm}_R$ self-consistently from the Schwinger-Dyson equation
  that corresponds to the diagrams in Fig. [7]. This expression for
$\Gamma^{lm}_R$ can then be  substituted into the integral
equation~(\ref{int2}) from which we obtain the viscosity. The diagrams in Fig.
[7] give,
\bea
&& \Gamma^{lm}_{abc}(K,Q, -K-Q) \nonumber \\
&&=  \int dP\,dR\,\tau_3^b \tau_3^a\tau_3^c I_{lm}(p,p) D_{ca}(R) D_{ac}(K+R-P)
D_{bc}(P+Q) D_{ab}(P) \\
&&+ \frac{1}{2} \int dP\,dR\, \Gamma^{lm}_{dbe}(P,Q,-P-Q)  D_{ad}(P)
D_{ec}(P+Q) D_{ac}(R+K-P) \tau_3^c D_{ca}(R) \tau_3^a \nonumber
\eea
We expand this expression by using the decompositions~(\ref{decompD1})
and~(\ref{decompver}) and performing the contractions. In the limit
$Q\rightarrow 0$ we make use of the pinch limit to obtain,
\bea
&&\label{gammaint1st}\Gamma^{lm}_R(K,Q,-K-Q)
=
-\frac{\lambda^2}{4}\int dP\,dR\,(I_{lm}(p,p)
+\frac{1}{2}\Gamma^{lm}_R(P,Q,-P-Q))\\
&&~~~~[f_p r_{p+q}(f'r_r + a' f_r)  + a_p r_{p+q}(r'a_r + a'r_r + f'f_r) + a_p
f_{p+q}(f'a_r + r'f_r)] \nonumber
\eea
where we have defined $r'=D_R(P');~P'=R+K-P$, etc. The key point is that this
integral equation is decoupled: there are no vertex components other than
$\Gamma^{lm}_R$ on the right hand side.
We introduce the notation,
\bea
\delta(r,k) &&= a_r r_{k} + r_r a_{k} + f_r f_{k} \nonumber \\
\phi_1(r,k)  && = f_r a_{k} + r_r f_{k} \label{delta-phi} \\
\phi_2(r,k)  &&= f_r r_{k} + a_r f_{k} \nonumber
\eea
which allows us to write,
\bea
\label{gammaint}
&&\Gamma^{lm}_R(K,Q,-K-Q)
=
-\frac{\lambda^2}{4}\int dP\,dR\,(I_{lm}(p,p)
+\frac{1}{2}\Gamma^{lm}_R(P,Q,-P-Q)) \\&&~~~~[f_p r_{p+q}\phi_1(r,r+k-p) + a_p
r_{p+q}\delta(r,r+k-p) + a_p f_{p+q}\phi_2(r,r+k-p)] \nonumber
\eea
In conclusion then, we can obtain a resummation of ladder graphs by
solving~(\ref{gammaint}) for $\Gamma^{lm}_R$ and substituting into the
expression (\ref{int2}) from which we obtain the viscosity.

\subsubsection{The Infrared Divergence}

It is easy to see that when the limit $Q\rightarrow 0$ is taken, the
resummation of ladder graphs is divergent. In fact, this result can be seen
immediately from~(\ref{int2}): the difference $\sigma-\sigma'$ is of order
$Q^0$ as $Q\rightarrow 0$ and therefore the viscosity, as obtained
from~(\ref{kubo}) and ~(\ref{sigma}), is divergent.
This problem is customarily remedied by replacing the bare propagators on the
rails (the horizontal lines) with hard thermal loop (HTL) effective propagators
\cite{Pisarski,gpy,par}. We write the HTL propagators as
\bea
D^*(P)=\frac{1}{P^2 - \Sigma(P)} \label{propstar}
\eea
where $\Sigma(P)$ is the HTL contribution to the self energy. Note that the
superscript $^*$ indicates a corrected propagator (and not complex
conjugation).
It is easy to see that this replacement regulates the divergence, but does not
invalidate the conclusion that all of the ladder graphs need to be resummed.
When resummed propagators are used, products of propagators of the form
$r^*_{k+q} a^*_k$ and $r^*_k a^*_{k+q}$ contribute factors on the order of one
over the imaginary part of the HTL self energy (instead of the factors $\sim
1/Q$ that occur for bare propagators). The imaginary part of the HTL self
energy is of order $\lambda^2$, and such a factor arises for each pair of
rails. These factors will compensate  the extra factors of $\lambda^2$  in the
numerator that accompany the addition of each rung.  Since each additional
rung is accompanied by an additional pair of rails, the result is that all of
the ladder graphs have a piece that contributes to leading order.

 To resum ladder diagrams with HTL effective propagators, we
rewrite~(\ref{int2}) with HTL propagators,
\bea
&&\sigma-\sigma'\label{int3}\\
&& = - \int  dK\, (N_k - N_{k+q}) [(I_{lm}(k,k)  +\frac{1}{2}
\Gamma^{lm}_R(K,Q,-K-Q)) a^*_k r^*_{k+q} + h.c.]I_{lm}(k,k)
\nonumber
\eea
and replace the vertex with the solution to the integral equation
(compare~(\ref{gammaint})),
\bea
\label{GGammaHTL}
&&\Gamma^{lm}_R(K,Q,-K-Q)
=
-\frac{\lambda^2}{4}\int dP\,dR\,(I_{lm}(p,p)
+\frac{1}{2}\Gamma^{lm}_R(P,Q,-P-Q)) \\&&~~~~[f^*_p r^*_{p+q}\phi^*_1(r,r+k-p)
+ a^*_p r^*_{p+q}\delta^*(r,r+k-p) + a^*_p f^*_{p+q}\phi^*_2(r,r+k-p)]
\nonumber
\eea
where $\phi^*_1$, $\phi_2^*$, and $\delta^*$ are obtained from
(\ref{delta-phi}) with bare propagators replaced by HTL propagators.
When the solution to~(\ref{GGammaHTL}) is substituted into~(\ref{int3}) we find
that $\sigma-\sigma'$ is of order $Q$, and the viscosity, as obtained
from~(\ref{kubo}) and ~(\ref{sigma}) is finite.

\subsubsection{The Boltzmann Equation}

If we use the HTL self energy to
two loop order and include a thermal mass and a thermal width, these results
agree with results obtained previously in the imaginary time formalism
\cite{jeon2}.
In order to sum all planar ladder diagrams, Jeon has constructed a linear
integral
equation involving a $4\times 4$ matrix valued kernel for the effective vertex,
which is a four component column vector. The four components represent four
different ``cuts'' in the imaginary time formalism. In the limit $Q\rightarrow
0$, the $4\times4$ matrix valued kernel reduces to a rank one matrix and the
original matrix integral
equation reduces to a decoupled single component equation for the reduced
vertex,
which is a complicated  linear combination of the four different cuts of the
vertex. We will show below that this result is considerably more
straightforward to derive in the real time formalism.

We define a new vertex
\bea
\bar \Gamma_{lm} = I_{lm} + \frac{1}{2} \Gamma_{lm} \nonumber
\eea
In terms of this definition (\ref{int3}) becomes,
\bea
&&\sigma-\sigma'\label{int3p} \simeq - \int  dK\, (N_k - N_{k+q}) [\bar
\Gamma^{lm}_R(K,Q,-K-Q) a^*_k r^*_{k+q} + h.c.]I_{lm}(k,k)
\eea
The integral equation for $\bar \Gamma_{lm}$ is obtained by adding $I_{lm}$ to
both sides of one half of (\ref{GGammaHTL}) to obtain,
\bea
\label{Gammabar}
&&\bar \Gamma^{lm}_R(K,Q,-K-Q)
= I_{lm}(k,k)
-\frac{\lambda^2}{8}\int dP\,dR\,\bar \Gamma^{lm}_R(P,Q,-P-Q) \\&&~~~~[f^*_p
r^*_{p+q} (f^*_r {a'}^* + r^*_r {f'}^*) + a^*_p r^*_{p+q}(a^*_r {r'}^* + r^*_r
{a'}^* + f^*_r {f'}^*) + a^*_p f^*_{p+q} (f^*_r {r'}^* + a^*_r {f'}^*)]
\nonumber
\eea
where we have used the definitions (\ref{delta-phi}).

We define the spectral density function as,
\bea
\rho_p& =&i(r^*_p-a^*_p)\label{rho}
\eea
In the limit that $Q$ is small the pinching terms dominate and we can  rewrite
the second term in (\ref{Gammabar}) as,
\bea
\label{second}
-\frac{\lambda^2}{8}\int && dP\,dR\, \bar\Gamma^{lm}_R(P,Q,-P-Q)\nonumber
\\
&&a_p^* r_{p+q}^*[( r'^* a^*_r + a'^* r^*_r +f^*_r f'^*) - N_p(f^*_r a'^* +
r^*_r f'^*) + N_{p+q}(f^*_r r'^* + a^*_r f'^*)]
\eea
Using (\ref{propstar}) and (\ref{rho}) we have,
\bea
r^*_p a^*_p = - \frac{\rho_p}{2{\rm Im} \Sigma_R(P)}\label{ra}
\eea
and $f_p = -iN_p \rho_p$.
 Taking $Q\rightarrow 0$ and using these results we find that (\ref{second})
becomes,
\bea
- \frac{\lambda^2}{4}&&\int dP\,dR\, \frac{\bar\Gamma^{lm}_R(P)}{{\rm
Im}\Sigma_R(P)} \rho_{p}\rho_r\rho_{p'}[ n_r n_{p'}+ n_r +n_p (n_r-n_{p'})]
\eea
Using the identity
\bea
 n_r n_{p'}+ n_r +n_p (n_r-n_{p'}) = (1+n_p)(1+n_{p'})n_r / (1+n_k)
\eea
and substituting into (\ref{Gammabar}) we have,
\bea
\label{gammaintsh1}
\bar\Gamma_R^{lm}(K) = &&k_m k_l -\frac{1}{3}
 \delta_{ml} k^2 - \frac{\lambda^2}{4}\int dP\,dR\,dP'\,\nonumber \\
&& (2\pi)^4 \delta^4(P+P'-R-K)
\rho_p \rho_r\rho_{p'} \frac{\bar\Gamma_R^{lm}(P)}{{\rm Im}\Sigma_R(P)}
(1+n_p)(1+n_{p'})n_r/(1+n_k)
\eea
The appearance of the imaginary part of the HTL self energy in the denominator
of this result indicates that a finite width is necessary to regulate an
infrared divergence in the integral. Once the integral has been expressed in a
regulated form, the imaginary part of the self energy can be dropped in the
expressions for the spectral densities, since it is subleading. In
(\ref{gammaintsh1}) we take,
\bea
\rho_p \rightarrow \rho^0_p = 2\pi {\rm sgn}(p_0)\delta(P^2-m_{th}^2)
\label{rho0}
\eea
where $m^2_{th} := m^2 + {\rm Re}\Sigma_R$.
 We define
\bea
\bar{\Gamma}_R^{lm} &&:= \hat I_{lm}(k,k) \Gamma_R^{shear}
\label{shear}\\
\hat I_{lm}(k,k) &&:= \frac{1}{k^2} I_{lm}(k,k) = (\hat k _l \hat k_m -
\frac{1}{3} \delta_{lm})\nonumber \\
B(K) &&:= \frac{\Gamma_R^{shear}(K)}{{\rm Im}\Sigma_R(K)}\nonumber,
\eea
and symmetrize to obtain from (\ref{gammaintsh1}),
\bea
\label{gammaintsh11}
\bar\Gamma_R^{lm}(K) = && I_{lm}(k,k)  - \frac{\lambda^2}{4}\int dP\,dR\,dP'\,
(2\pi)^4 \delta^4(P+P'-R-K) \rho^0_p \rho^0_r\rho^0_{p'}\nonumber \\
&&
 \frac{1}{3}(\hat I_{lm}(p)B_p + \hat I_{lm}(p')B_{p'} - \hat I_{lm}(r)B_r)
(1+n_p)(1+n_{p'})n_r/(1+n_k) \nonumber
\eea
Rearranging we obtain,
\bea
I_{lm}(k,k) = && {\rm Im}\Sigma_R( K) B(K) \hat I_{lm}(k,k) +
\frac{\lambda^2}{12(1+n_k)} \int dP\,dR\,dP'~(2\pi)^4 \delta({P} + {P'} - {R} -
{K})\nonumber \\
&& \rho^0_p \rho^0_r\rho^0_{p'} [\hat I_{lm}(p,p) B({P}) + \hat I_{lm}(p',p')
B({P'}) - \hat I_{lm}(r,r) B({R})] (1+n_p)(1+n_{p'})n_r\label{gammanew}
\eea
%Next we do the integrals over the zeroth momentum components. In order to
%%%%requencies (IS THIS OKAY)?)  We obtain,
%\bea
%\bar\Gamma_R^{lm}(\underline{K}) =&& k_m k_l -\frac{1}{3}
% \delta_{ml} k^2 -\frac{\lambda^2}{6(1+n(\omega_k)} \int \tilde d\,p ~\tilde d
%%%\,r ~\tilde d \,p'~(2\pi)^4 \delta(\underline{P} + \underline{P'} -
%%%\underline{R} - \underline{K}) \nonumber \\
%&& [\hat I(p)^{lm} B(\underline{P}) +\hat I(p')^{lm} B(\underline{P'})  - \hat
%%%I(r)^{lm} B(\underline{R})] (1+n(\omega_p))(1+n(\omega_{p'}))n(\omega_r)
%%%\nonumber
%\eea
%where $\tilde d\,p = \frac{d^3p}{(2\pi)^3 2\omega_p}$, $\omega_p^2 = p^2 +
%%%m_0^2 + {\rm Re}\Sigma_R$ and a momentum variable that is underlined
%%indicates %an onshell mommentum:  $\underline{K} = (\omega_k, \vec{k})$.
%%Rearranging, we %obtain,
%\bea
% \hat I(k)^{lm}  =&& {\rm Im}\Sigma_R(\underline K) B(\underline K) +
%%%\frac{\lambda^2}{6(1+n(\omega_k))} \int \tilde d\,p ~\tilde d \,r ~\tilde d
%%%\,p'~(2\pi)^4 \delta(\underline{P} + \underline{P'} - \underline{R} -
%%%\underline{K})  \\
%&& [\hat I(p)^{lm} B(\underline{P}) + \hat I(p')^{lm} B(\underline{P'}) - \hat
%%%I(r)^{lm} B(\underline{R})]
%%%(1+n(\omega_p))(1+n(\omega_{p'}))n(\omega_r)\label{gammashell}
%\eea
In $\phi^4$ theory ${\rm Im}\Sigma_R$  can be
expressed as,
\bea
{\rm Im}\Sigma_R(K) =&&-\frac{\lambda^2}{12}\left(\frac{1}{(1+n_k)}\right) \int
dP\,dR\ dP'\, (2\pi)^4
             \delta^4(P+P'-R-K) \nonumber \\
&&\rho^0_p\rho^0_r\rho^0_{p'}(1+n_p)(1+n_{p'})n_r\label{pi}
%=&&-\frac{1}{12}\frac{1}{(1+n(\omega_k))} \lambda^2 \int \tilde d\,p ~\tilde d
%%%\,r ~\tilde d \,p'~ (2\pi)^4 \delta(\underline{P} + \underline{P'} -
%%%\underline{R} - \underline{K})
% \nonumber \\
%&& (1+n(\omega_p))(1+n(\omega_{p'}))n(\omega_r).
\eea
which allows us to rewrite (\ref{gammanew}) as,
\bea
&& I(k,k)_{lm} = \frac{\lambda^2}{12(1+n_k)} \int dP\,dR\,dP' ~(2\pi)^4
\delta({P} + {P'} - {R} - {K}) \rho^0_p \rho^0_r\rho^0_{p'}\label{Bint} \\
&& ~~~~ [\hat I_{lm}(p,p) B({P}) + \hat I_{lm}(p',p') B({P'}) - \hat
I_{lm}(k,k) B({K}) - \hat I_{lm}(r,r) B({R})] (1+n_p)(1+n_{p'})n_r
\nonumber
\eea

Equation (\ref{Bint}) is an integral equation whose solution represents a
non-perturbative resummation of contributions to $B$. The last step, is to
obtain an expression for the viscosity in terms of the quantity $B$. Using
(\ref{kubo}), (\ref{etap}), (\ref{int3p}) and (\ref{shear}) we have,
\bea
\eta = \frac{\beta}{30}  \lim_{q_0 \to 0}\lim_{\vec{q} \to 0} (1+n_q)  \int dK
(N_k - N_{k+q}) k^2 [ \Gamma_R^{shear}(K,Q,-K-Q) a_k^* r_{k+q}^* + h.c.]
\eea
Using
\bea
\lim_{q_0 \to 0}(1+n_q)(N_k-N_{k+q}) = 2n_k(1+n_k)
\eea
we obtain,
\bea
\eta = \frac{\beta}{15} \int dK\, k^2 n_k(1+n_k) [\Gamma_R^{shear} a_k^*
r_{k}^* + h.c.]
\eea
Using (\ref{ra}) and
(\ref{shear}) and making the substitution $\rho_p \rightarrow \rho^0_p$  we have,
\bea
\eta = \frac{\beta}{15} \int dK\,k^2 n_k(1+n_k) B_k \rho^0_k \label{BV1}
\eea
%After using the delta function (\ref{rho0}) to do the frequency integral we
%%obtain,
%\bea
%\eta = \frac{\beta}{30}  \int \tilde d k k^2 n(\omega_k) (1+n(\omega_k))
%%%B(\underline{K})
%\label{BV}
%\eea

It is straightforward to show that our integral equation for the resummation of
ladder contributions to the three point vertex (\ref{Bint}) has exactly the
same structure as the equation obtained from the linearized Boltzmann equation,
and that our expression for the shear viscosity (\ref{BV1}) is exactly the same
as that obtained by defining the shear viscosity as a transport coefficient and
using the linearized Boltzmann equation to get an integral equation for
viscosity.
We begin from the Boltzmann equation for the phase space distribution function
$f(\underline{X},\underline{P})$ which describes the evolution of the phase
space probability density for the fundamental particles comprising a fluid,
\bea
\frac{\underline{K}^\mu}{\omega_k} \frac{\partial}{\partial X^\mu}
f(X,\underline{K}) = \frac{1}{2} \int_{123} d \,\Gamma_{12\leftrightarrow 3k}[
f_1 f_2 (1+f_3)(1+f_k) - (1+f_1)(1+f_2) f_3 f_k ]
\eea
where $f_i:= f(X,\underline{K}_i)$, $f_k:=f(X,\underline{K})$ and $d \, \Gamma
_{12\leftrightarrow 3k}$ is the differential transition rate for particles of
momentum $P_1$ and $P_2$ to scatter into momenta $P_3$ and $K$ and is given by
\bea
d\,\Gamma_{12\leftrightarrow 3k} := \frac{1}{2\omega_k} |{\cal
T}(\underline{K},\underline{P}_3;\underline{P}_2,\underline{P}_1)|^2
\Pi^3_{i=1} \frac{d^3  p_i}{(2\pi)^32\omega_{p_i}} (2\pi)^4
\delta(\underline{P}_1 + \underline{P}_2 - \underline{P}_3 - \underline{K})
\eea
${\cal T}$ is a multiparticle scattering amplitude describing a process in
which particles of momenta $P_1$ and $P_2$ scatter into momenta $P_3$ and $K$.
Underlined momentum variables are on shell, and for the purposes of this
section we specialize further to the positive mass shell. The Boltzmann
equation is valid for distribution functions that describe the distributions of
positive energy particles.

In a weak coupling $\phi^4$ theory, one can linearize the
Boltzmann equation by expanding the non-equilibrium distribution function
$ f(X,\underline{P})$ around a local equilibrium function
$f_0(X,\underline{P})$:
\begin{equation}
f(X,\underline{P})=f_0(X,\underline{P})\{1-\chi(X,\underline{P})[1+f_0(X,\underline{P})]\}
\end{equation}
where
\bea
f_0(X,\underline{P}) = n(|u^\mu \underline{P_\mu})|\,;~~~~n(\omega) =
\frac{1}{e^{\beta \omega} -1}
\eea
Expanding in powers of $\nabla u$ one finds \cite{jeon2},
\begin{equation}
\chi (X,\underline{P})=\beta(X)A(X,\underline{P})\nabla\cdot
u(X)+\beta(X)B(X,\underline{P})[\hat{\underline{P}}
\cdot \nabla (u(X)\cdot \hat{\underline{P}})-\frac{1}{3}\nabla \cdot u(X)]
\end{equation}
The coefficient $A$ multiplying the divergence of the flow is related to the
bulk viscosity. We obtain the shear viscosity from the coefficient $B$ which
multiples the
shear in the flow. $B$ satisfies the linear inhomogeneous
integral equation\cite{jeon1}:
\begin{eqnarray}
 k_lk_m &&{-} {\textstyle{1\over 3}}\delta_{lm} {k}^2
          \displaystyle =  \displaystyle
           {1\over 4}\,\int
            \prod_{i=1}^3 {d^3 {p}_i \over 2E_{i}(2\pi)^3}\,\,
             (2\pi)^4 \delta(\underline{P}_1 {+} \underline{P}_2
             {-} \underline{P}_3 {-} \underline{K})\,
             |{\cal T} (\underline{P}_1,\underline{P}_2;
             \underline{P}_3,\underline{K})|^2\,
 \nonumber\\
        & \displaystyle & \displaystyle \qquad {}\times
               [1{+}n(\omega_1)]\, [1{+}n(\omega_2)]\,
               n(\omega_3)/ [1{+}n(\omega_k)] \,
\label{kinetic} \\
        & \displaystyle & \displaystyle \qquad {}\times
              ( \hat I_{lm}(k,k) B(\underline{K}) + \hat
I_{lm}(p_3,p_3)B(\underline{P_3})
              - \hat I_{lm}(p_2,p_2)B(\underline{P_2})- \hat
I_{lm}(p_1,p_1)B(\underline{P_1}) )
 \;,
 \nonumber
 \end{eqnarray}
We want to compare this result with (\ref{Bint}). The two equations have
exactly the same form once the delta functions have been used to do the
frequency integrals in (\ref{Bint}) and the classical piece, which corresponds
to the positive mass shell, is extracted.  The shear viscosity can be written
in terms of $B$ by looking at first order corrections to the energy momentum
tensor, and comparing with the constitutive relation,
\bea
\langle T_{ij} \rangle \simeq - \frac{\eta}{\langle \epsilon + {\cal P}\rangle}
[\nabla_i\langle T^0_j\rangle + \nabla_j\langle T^0_i\rangle - \frac{2}{3}
\delta_{ij} \nabla^l\langle T^0_l\rangle ] - \frac{\zeta}{\langle \epsilon +
{\cal P}\rangle} \delta_{ij} \nabla^l\langle T^0_l\rangle + \delta_{ij} \langle
{\cal P}\rangle
\eea
where $\epsilon := T_{00}$ is the energy density, ${\cal P} = \frac{1}{3}T^i_i$
is the pressure, and $\eta$ and $\zeta$ are the shear and bulk viscosities
respectively. The result is \cite{jeon1,jeon2}
\bea
\eta = \frac{\beta}{15}\int\frac{d^3k}{(2\pi)^3 \omega_k} k^2
n(\omega_k)(1+n(\omega_k))B(\underline{K})
\eea
We want to compare this expression with our result (\ref{BV1}). After doing the
integral over the frequency components in (\ref{BV1}) and extracting the
classical piece, the two expressions agree.

\subsection{Ladders from four-point vertices}
We can also resum ladder contributions to the viscosity by starting from the
integral equation for viscosity in terms of the four-point
function~(\ref{intM2}) and replacing the four-point function by the solution of
the integral equation shown in Fig. [8].  The zeroth order terms come from the
first graph on the right hand side of the figure. We obtain the term
corresponding to the second graph in the figure by using the
decomposition~(\ref{decompM}). We have,
\bea
&&M_B =  \frac{\lambda^2}{4}\int dR\, \delta(r,r+k-p) \nonumber \\
&&~~~~~-\frac{\lambda^2 }{32} \int  dR\,dS \left[ M_B \delta(r,r+s-p) a_s
r_{s+q}\right.
+ M_{R3}(\delta(r,r+s-p) a_s f_{s+q} \nonumber \\
&&~~~~~~+ \phi_2(r,r+s-p) a_s a_{s+q}) + M_{R1} (\delta(r,r+s-p) f_s r_{s+q} +
\phi_1(r,r+s-p) r_s r_{s+q})\left.\right] \nonumber \\
&&M_{R1} = \frac{\lambda^2}{4}\int dR\, \phi_1(r,r+k-p) \nonumber \\
&&~~~~~-\frac{\lambda^2 }{32} \int dR\,dS\,\left[ \right.
M_{R1}(\delta(r,r+s-p) r_s r_{s+q}
+ \phi_1(r,r+s-p) f_s r_{s+q}) \nonumber \\
&& ~~~~~~+ M_{R3} \phi_1(r,r+s-p) a_s f_{s+q} + M_B \phi_1(r,r+s-p) a_s
r_{s+q})\left. \right] \label{Mladder} \\
&&M_{R3} = \frac{\lambda^2}{4}\int dR\, \phi_2(r,r+k-p) \nonumber \\
&&~~~~~ -\frac{\lambda^2 }{32} \int dR\,dS\,\left[ \right.
M_{R3}(\delta(r,r+s-p) a_{s+q} a_s + \phi_2(r,r+s-p) f_{s+q} a_s) \nonumber \\
&& ~~~~~~+ M_{R1} \phi_2(r,r+s-p) r_{s+q} f_s + M_B \phi_2(r,r+s-p) r_{s+q} a_s
\left. \right] \nonumber
\eea
We can rewrite this result by taking the pinch limit and using the
definition~(\ref{Mtilde}). We obtain,
\bea
&&\tilde{M}(P+Q,-K-Q,-P,K) = \tilde{M}^{(0)}(P+Q,-K-Q,-P,K) -\frac{\lambda^2
}{32} \int dR\,dS \,a_s r_{s+q} \label{Mladder2}\\
&&\,\tilde{M}(S+Q, -K-S, -P, K) [\delta(r,r+s-p)  -N_p\phi_1(r,r+s-p)+ N_{p+q}
\phi_2(r,r+s-p)]  \nonumber
\eea
Note that as with~(\ref{gammaint}) this integral equation is decoupled. As a
consequence, we can obtain a resummation of ladder graphs by
solving~(\ref{Mladder2}) for $\tilde{M}$ and substituting into the integral
equation (\ref{intM4b}) from which we obtain the viscosity.

\section{Another way to resum graphs}

In this section we will show that there is another way to obtain a resummation
of contributions to the viscosity. We will see that the method discussed in
this section indicates that the ladder graphs are not the only graphs that need
to be resummed, in spite of the arguments discussed in the previous section.  We will identify the extra graphs that are  included in our resummation, which are not part of the conventional ladder resummation.  
The idea is as follows. In \cite{meg1} it is shown that in $\phi^3$ theory,
the retarded three-point vertex can be related to the self energy by a relation
of the form,
\begin{equation}
\label{g-p}
\Gamma_R(K,Q,-K-Q) \simeq {1\over Q^2+ 2Q\cdot K} (\Pi_A(K)- \Pi_R(Q+K))
\end{equation}
where the symbol $\simeq$ indicates that we are looking at the infrared region
of the momentum integrals in the various terms in the loop expansion of the
three-point vertex and the self energy. This condition will be made explicit
below. We will obtain an equation of this form for the corrected three-point
vertices involved in~(\ref{int2}), and for the corrected four-point vertices
in~(\ref{intM4b}). In both cases, a resummation is obtained by replacing the
self energy in these expressions by the solution of a separate coupled integral
equation that resums an infinite series of contributions to the self energy.
This procedure is presented in detail below.

\subsection{The corrected three-point function}

  In~(\ref{int2}) the vertex~(\ref{GGamma}) is similar to the three-point
vertex in $\phi^3$ theory except that it contains the composite operator
$\pi_{lm}(X)$ ~(\ref{ppi}). We will show that in the limit of infrared loop momenta this
three-point function is related to a function that is similar to the sunset
self energy from $\phi^4$ theory. The equation that relates these quantities
is similar to~(\ref{g-p}). At one loop, the relevant diagrams are shown in Fig
[9]. In the vertex diagram, the place where the two lines join is a
bare vertex $\Gamma^{(0)lm}_{abc}$ ~(\ref{GGamma0}). In the self energy
diagram, a line carrying momentum $P$ with an asterix contains a factor
$I_{lm}(p,p)$. Note that this notation does not indicate an insertion, but merely a multiplicative factor. We will show explicitly that the diagrams in Fig [9] satisfy
the relation,
\begin{equation}
\label{g-p-lm}
\Gamma^{lm}_R(K,Q,-K-Q) \simeq {2\over Q^2+ 2Q\cdot K} (\Pi^{lm}_A(K)-
\Pi^{lm}_R(Q+K))
\end{equation}
Note that in order to obtain this result we have rewritten $I_{lm}(p+k,p+k+q) = I_{lm}(p+k,p+k) + I_{lm}(p+k,q)$ and taken only the first term.  We have not included the extra contributions that result from the second term, since these terms will not contribute to the final result when the limit $Q\rightarrow 0$ is taken.
We verify this relation explicitly at one loop, and we also discuss the relation at higher orders in the loop expansion, working at
zero temperature for simplicity, and discuss what sets of graphs are included.
 Finally, we will show that when this expression for the vertex is substituted
into the integral equation (\ref{int2}) from which we obtain the viscosity, and
the self energy $\Pi^{lm}$ is replaced by the solution of a separate integral
equation, the result for the viscosity corresponds to an infinite resummation
of graphs.

\subsubsection{The splitting relation for the three-point function}

We begin by deriving~(\ref{g-p-lm}). The three-point function in Fig. [9a] is
given by the expression
\begin{equation}
\Gamma^{lm}_{ebd
}=  \int dP\int dR(-i\lambda)^2 iD_{de}(R)\tau_3^e iD'_{ed}
\tau_3^d iD_{eb}(P) iD_{bd}(P+Q) \tau_3^b I_{lm}(p,p) \end{equation}
%Expanding this equation using the vector column representation of
%%propagators~(\ref{decompD1}) we obtain,
%\begin{eqnarray}
%\Gamma^{lm}_{ebd} =
%{-\lambda^2\over 2^4}\int dP \int dR
%&& \left(r_R{1\choose 1}{1\choose 1}
%+a_R{1\choose -1}{1\choose -1}+
%f_R{1\choose 1}{1\choose -1}\right)_{de}
%\nonumber\\
%&& \left(r_R'{1\choose 1}{1\choose 1}
%+a_R'{1\choose -1}{1\choose -1}+
%f_R'{1\choose 1}{1\choose -1}\right)_{ed}
%\nonumber\\
%&& \left(r_p{1\choose 1}{1\choose -1}
%+a_P{1\choose -1}{1\choose 1} +
%f_P{1\choose 1}{1\choose 1}\right)_{eb}
%\\
%&& \left(r_{p+q}{1\choose 1}{1\choose -1}
%+a_{p+q}{1\choose -1}{1\choose 1}+
%f_{p+q}{1\choose 1}{1\choose 1}\right)_{bd}\tau_3^b I_{lm}(P,P+Q) \nonumber
%\end{eqnarray}
We expand this equation using the vector column representation of the
propagators~(\ref{decompD1}) and extract $\Gamma^{lm}_R$~(\ref{eq: physVer2}).
We make the change of variable $P\rightarrow P+K$. We will need the functions
$\delta(r,r-p)$, $\phi_1(r,r-p)$ and  $\phi_2(r,r-p)$. These expressions are
defined in~(\ref{delta-phi}). In the equations that follow we will suppress the
momentum arguments for these quantities.
We obtain,
\begin{eqnarray}
\Gamma^{lm}_R(K, Q, -K-Q)=
&&-{\lambda^2\over 4}\int dP dR\,I_{lm}(p+k,p+k)\nonumber \\
&&~~~~~~( f_{p+k}r_{p+k+q}\phi_1 + f_{p+k+q}a_{p+k}\phi_2
+ r_{p+k+q}a_{p+k}\delta )\label{verunsplit}
\end{eqnarray}
Splitting the propagator pair and looking at the region of phase space where
\bea
2P\cdot Q \ll Q^2+2K\cdot Q
\label{cond1}
\eea
 we have,
\begin{equation}
\label{splitprop}
D(P+K+Q)D(P+K)\simeq {D(P+K)-D(P+K+Q)\over Q^2+2K\cdot Q}
 \end{equation}
which gives,
\begin{eqnarray}
\label{gamma_r}
\Gamma^{lm} _R(K, Q, -K-Q)=&&
-{\lambda^2\over 4 (Q^2+2Q\cdot K)} \int dP dR I_{lm}(p+k,p+k) \nonumber\\
&&~~~~~~[(a_{p+k}-r_{p+k+q})\delta + f_{p+k}\phi_1- f_{p+k+q}\phi_2 ]
\end{eqnarray}
We compare this result with the sunset self energy as shown in Fig. [9b]. We
obtain,
\begin{eqnarray}
\Pi_0(K)^{lm}_{ab}=
i\frac{1}{2} \int dR dP (i\lambda)^2 i^3 \tau_3^a \tau_3^b D_{ba}(R)
D_{ab}(P+K) I_{lm}(p+k,p+k)
D_{ab}(R-P)
%\nonumber\\
%&&=\frac{\lambda^2}{8}\int dR dP \left(r_R{1\choose 1}{1\choose -1}
%+a_R{1\choose -1}{1\choose 1}
%+f_R{1\choose 1}{1\choose 1}\right)_{ba}
%\nonumber\\
%&& ~~~~\left(r_{P+K}{1\choose -1}{1\choose 1}
%+a_{P+K}{1\choose 1}{1\choose -1}
%+f_{P+K}{1\choose -1}{1\choose -1}\right)_{ab}
%\\
%&& ~~~~\left(r_{R-P}{1\choose 1}{1\choose -1}
%+a_{R-P}{1\choose -1}{1\choose 1}
%+f_{R-P}{1\choose 1}{1\choose 1}\right)_{ab}I_{lm}(P+K,P+K) \nonumber
\end{eqnarray}
Contracting indices and using the definitions of retarded and advanced
self-energies~(\ref{physPi}),
we obtain,
\begin{eqnarray}
\label{pira}
\Pi_0(K)^{lm}_R &=&-{\lambda^2\over 8}\int dR dP I_{lm}(p+k,p+k)
( r_{p+k}\delta(r,r-p) + f_{p+k}\phi_2(r,r-p))\\
\Pi_0(K)^{lm}_A &=&-{\lambda^2\over 8}\int dR dP I_{lm}(p+k,p+k)
( a_{p+k}\delta(r,r-p)
+ f_{p+k}\phi_1(r,r-p)) \nonumber
\end{eqnarray}
Comparing (\ref{gamma_r}) with ({\ref{pira}) gives the result~(\ref{g-p-lm}).

When we look at~(\ref{g-p-lm}) at higher order in the loop expansion, we find
that the sets of graphs that satisfy this relation do not correspond to only
the ladder graphs. Working at zero temperature for simplicity, it is
straightforward to show that the vertex graphs shown in Fig. [10] correspond to
the second order sunset graph shown in Fig. [11]. The important point is that
there is a cancellation between contributions from the ladder graph (Fig.
[10a]) and contributions from the two non-ladder graphs (Fig [10b, 10c]). This
cancellation indicates that the leading contributions from the ladder graphs
must be of the same order as the leading contributions from the non-ladder
graphs, in contradiction to the prediction of the pinch argument, as described
in the previous section. As explained previously, the conclusion that
non-ladders are suppressed relative to ladders is based on the following
argument. There is only a limited region of phase space (the infrared region)
in which non-ladders are significant, and since the measure that corresponds to
this region is small, the non-ladders should be suppressed. However, our
derivation of~(\ref{g-p-lm}) is based on the realization that in the limit that
$Q\rightarrow 0$ it is precisely this region of phase space that is important,
since both ladders and non-ladders are largest in  this region. Since we have
singled out the infrared part of the phase space, by imposing the condition
$2P\cdot Q \ll Q^2 + 2K\cdot Q$, the argument about the measure being small is
no longer relevant and there are contributions from the non-ladder graphs that
are as large as contributions from the ladder graphs.

\subsubsection{An Integral equation for $\Pi$}

The second step in this procedure is to obtain an integral equation that
resums an infinite set of contributions to $\Pi$. The integral equation
that resums the sunset contributions to the self energy is shown in Fig. [12].
We obtain,
\bea
\Pi(K)^{lm}_{ab} = &&\Pi_0(K)_{ab}^{lm} \nonumber\\
&& -\frac{\lambda^2}{2}\int dR\,dP \tau^3_a \tau^3_b D_{ba}(R) D_{ab}(R-P)
D_{ac}(K+P) \Pi^{lm}_{cd}(K+P) D_{db}(K+P)
\eea
Contracting indices we obtain, for the retarded and advanced combinations,
\bea
\Pi(K)_R^{lm} = &&\Pi_0(K)_{R}^{lm} -\frac{\lambda^2 }{8}\int dR\,dP\,\left(
\delta [\Pi(K+P)_R^{lm} r_{k+p} r_{k+p}]\right. \nonumber
\\
&&+\left. \phi_2[\Pi(K+P)_R^{lm} f_{k+p} r_{k+p} + \Pi(K+P)_A^{lm} a_{k+p}
f_{k+p} + \Pi(K+P)_F^{lm} a_{k+p} r_{k+p}]\right) \nonumber \\
\Pi(K)_A^{lm} =  &&\Pi_0(K)_{A}^{lm} -\frac{\lambda^2 }{8}\int dR\,dP\,\left(
\delta [\Pi(K+P)_A^{lm} a_{k+p} a_{k+p}] \right.\label{pi-int} \\
&& + \left.\phi_1[\Pi(K+P)_R^{lm} f_{k+p} r_{k+p} + \Pi(K+P)_A^{lm} a_{k+p}
f_{k+p} + \Pi(K+P)_F^{lm} a_{k+p} r_{k+p}]\right) \nonumber
\eea
where $\delta,~~\phi_1, ~~\phi_2$ are functions $f(r,r-p)$ and the momentum
dependence has been suppressed.

One of the advantages of the resummation technique outlined in this section is
that it can be easily extended to include other types of graphs. By using an
integral equation that resums different contributions to the self energy, one
can obtain the resummation of a different set of contributions to the
viscosity.

\subsubsection{Contributions to the Viscosity}

Finally, we ask what contributions to the viscosity are obtained by
substituting the solution of the integral equation~(\ref{pi-int})
into~(\ref{g-p-lm}) (which is in turn substituted into~(\ref{int2})). The
first term in~(\ref{int2}) is the one loop graph. We will show that the second
term is the first ladder contribution to the viscosity (Fig. [13a]), when we
use~(\ref{g-p-lm}) with the self energy given by the first order sunset
graph~(\ref{pira}).  Substituting~(\ref{g-p-lm}) and ~(\ref{pira})
into~(\ref{int2}) we obtain for the second term,
\bea
(\sigma - \sigma')^{(2)} = -\frac{\lambda^2}{8} \int dK\,&&\frac{1}{Q^2 +
2Q\cdot K} \int dR\,dP\, I_{lm}(p+k,p+k) \{(N_k - N_{k+q})a_k r_{k+q} \nonumber
\\
&&[(a_{p+k} - r_{p+k+q})\delta + f_{p+k}\phi_1 -  f_{p+k+q}\phi_2] +
h.c.\}I_{lm}(k,k)
\eea
As before, $\delta,~~\phi_1,~~\phi_2$ are functions $f(r,r-p)$ and the momentum
dependence has been suppressed.
We use~(\ref{splitprop}) to obtain,
\bea
\label{comp1}
(\sigma - \sigma')^{(2)} = -\frac{\lambda^2}{8}\int
dK\,dR\,dP\,&&I_{lm}(p+k,p+k) \{ (N_k - N_{k+q})a_k r_{k+q} ( \delta a_{p+k}
r_{p+k+q} \nonumber \\
&&+ \phi_1 f_{p+k} r_{p+k+q} + \phi_2 a_{p+k} f_{p+k+q}) + h.c. \}I_{lm}(k,k)
\eea
We want to compare this result with the result obtained from the first ladder
graph shown in Fig. [13a]. Defining the quantity in the figure as
$\lambda_{ab}$ we want to look at $\sigma - \sigma' = \lambda_{21} -
\lambda_{12}$. From the figure we have,
\bea
\lambda_{ab} = \lambda^2\int dR\,dP \tau_3^c \tau_3^d && D_{cd}(R) D_{dc}(R-P)
I_{lm}(p+k,p+k) \nonumber \\
&&D_{da}(P+K) D_{ac}(P+K+Q) D_{cb}(K+Q) D_{bd}(K) I_{lm}(k,k)\nonumber
\eea
We substitute in the vector forms for the propagators~(\ref{decompD1}). We
keep only the terms that give us non-zero contributions to
$(\sigma-\sigma')^{(2)}$. The result is,
\bea
\label{comp2}
(\sigma-\sigma')^{(2)} = -\frac{\lambda^2}{8} \int
dK\,dR\,dP\,&&I_{lm}(p+k,p+k) \{ (N_k - N_{k+q})a_k r_{k+q} (\delta a_{p+k}
r_{p+k+q} +\nonumber \\
&&  \phi_1 f_{p+k} r_{p+k+q} + \phi_2 a_{p+k} f_{p+k+q} ) + h.c. \} I_{lm}(k,k)
\eea
where we have made the change of variable $K \rightarrow -K-P-Q$ in the second
term in order to obtain a form that is easily recognizable as the hermitian
conjugate. Once again, $\delta,~~\phi_1~~\phi_2$ are functions $f(r,r-p)$ and
the momentum dependence has been suppressed. Comparison of~(\ref{comp1})
and~(\ref{comp2}) verifies our statement that the first order sunset graph
corresponds to the first order ladder graph.

At higher orders the calculation is more complicated. In Appendix A, we
discuss the two loop contributions, working at zero temperature for simplicity.
 We show that the second order calculation gives contributions to the viscosity
of the form shown in Fig. [13b] and [13c]. The key point is that non-ladder
graphs are included.
The conclusion is that we can obtain a resummation of ladder and non-ladder
graphs by substituting the solution of the integral equation that resums sunset
graphs~(\ref{pi-int}) into the expression for the vertex~(\ref{g-p}), which is
in turn substituted into the integral equation (\ref{int2}) from which we
obtain the viscosity.

\subsubsection{The Infrared Divergence}

The procedure described above will result in an expression that is divergent in
the limit $Q\rightarrow 0$, in exactly the same way that the previous result,
obtained by solving the integral equation for the vertex~(\ref{gammaint}), was
divergent. In this case, we expect to encounter difficulties in the limit
$Q\rightarrow 0$ since the condition (\ref{cond1}) is not satisfied in this
limit.
We can remedy the problem by using HTL propagators. The vertex is the same as
the vertex shown in Fig. [9a], except that the propagators are now HTL
propagators~(\ref{propstar}). From~(\ref{verunsplit}) we obtain,
\begin{eqnarray}
\tilde{\Gamma}^{lm}_R(K, Q, K-Q)=
-{\lambda^2\over 4}\int dP dR\,&&I_{lm}(p+k,p+k)\nonumber \\
&&( f^*_{p+k}r^*_{p+k+q}\phi^*_1 + f^*_{p+k+q}a^*_{p+k}\phi^*_2
+ r^*_{p+k+q}a^*_{p+k}\delta^*)
\end{eqnarray}
We split propagator pairs as before. With HTL propagators, the splitting
relation has the form,
\begin{equation}
D_R^*(P+K+Q)D_A^*(P+K)\simeq {D_A^*(P+K)-D_R^*(P+K+Q)\over Q^2+2K\cdot Q -
\Sigma_R(K+Q) + \Sigma_A(K)}
 \end{equation}
In this case, the notation $\simeq$ indicates that we are looking at the region
of phase space where
\bea
2Q\cdot P +P(\Sigma'_A(K) - \Sigma_R'(K+Q)) \ll Q^2 + 2Q\cdot K - \Sigma_R(K+Q)
+ \Sigma_A(K) \label{cond2}
\eea
This condition can be compared to the condition for the splitting of bare
propagators (\ref{cond1}). Since the imaginary part of the HTL sunset self
energy is non-zero, the condition (\ref{cond2}) is satisfied in the infrared
region of the momentum integral, even when $Q$ is taken to zero.
The resulting expression has the form,
\bea
\label{gammar2}
&&\Gamma^{lm}_R(K,Q,-K-Q) = -\frac{\lambda^2}{4} \left( \frac{1}{Q^2 + 2Q\cdot
K + \Sigma_A(K) - \Sigma_R(K+Q)}\right) \nonumber \\
&&~~~~\int dP\, dR\, I_{lm}(p+k,p+k) [(a^*_{p+k} - r^*_{p+k+q})\delta^* +
f^*_{p+k}\phi^*_1 - f^*_{p+k+q}\phi^*_2 ]
\eea
We compare this result with the sunset self energy (compare (\ref{pira})),
\begin{eqnarray}
\label{pira2}
\Pi_0(K)^{lm}_R &=&-{\lambda^2\over 8}\int dR dP\,I_{lm}(p+k,p+k)
( r^*_{p+k}\delta^*
+ f^*_{p+k}\phi^*_2)
\\
\Pi_0(K)^{lm}_A &=&-{\lambda^2\over 8}\int dR dP\,I_{lm}(p+k,p+k)
( a^*_{p+k}\delta^*
+ f^*_{p+k}\phi^*_1) \nonumber
\end{eqnarray}
where $\delta,~~\phi_1,~~\phi_2$ are functions $f(r,r-p)$ and the momentum
dependence has been suppressed.
Comparing~(\ref{gammar2}) and~(\ref{pira2}) we obtain,
\bea
\label{g-p2}
\Gamma^{lm}_R(K,Q,-K-Q)={2\over Q^2+ 2Q\cdot K -\Sigma_R(Q+K) + \Sigma_A(K)}
(\Pi^{lm}_A(K)- \Pi^{lm}_R(Q+K))
\eea

We obtain the self energy from the integral equation (compare~(\ref{pi-int})),
\bea
\Pi(K)_R^{lm} =&&\Pi_0(K)_{R}^{lm}  -\frac{\lambda^2}{8}\int dR\,dP\,\left(
\delta^* [\Pi(K+P)_R^{lm} r^*_{k+p} r^*_{k+p}]\right. \nonumber
\\
&&+\left. \phi^*_2[\Pi(K+P)_R^{lm} f^*_{k+p} r^*_{k+p} + \Pi(K+P)_A^{lm}
a^*_{k+p} f^*_{k+p} + \Pi(K+P)_F^{lm} a^*_{k+p} r^*_{k+p}]\right) \nonumber \\
\Pi(K)_A^{lm} =  &&\Pi_0(K)_{A}^{lm} -\frac{\lambda^2}{8}\int dR\,dP\,\left(
\delta^* [\Pi(K+P)_A^{lm} a^*_{k+p} a^*_{k+p}] \right.\label{pi-int2} \\
&& + \left.\phi^*_1[\Pi(K+P)_R^{lm} f^*_{k+p} r^*_{k+p} + \Pi(K+P)_A^{lm}
a^*_{k+p} f^*_{k+p} + \Pi(K+P)_F^{lm} a^*_{k+p} r^*_{k+p}]\right) \nonumber
\eea
The corresponding contributions to the viscosity are the same as in Fig. [13], but with all lines taken as HTL propagators.  Note that, as usual when calculating with HTL propagators, we must include counter terms to avoid double counting.  

\subsection{The corrected four-point function}

It is also possible to obtain an expression similar to~(\ref{g-p}) for the
four-point function. We must look at the special case where two of the four
fields have the same position co-ordinate, and the same Keldysh index. A
vertex of this form is shown in Fig. [14]. Note that this restriction means
that the 1PI four-point vertex has essentially the same form as a three-point
vertex. We will show that the use of this vertex in the expression for the
viscosity~(\ref{intM2}) leads to a resummation of contributions to the
viscosity, all of which contain one chain link. A diagram of this form is
shown in Fig. [15]. It has been shown  that diagrams containing chain links
are suppressed \cite{jeon1}. The argument is as follows. As discussed in
section [V-A-1] in the case of ladder diagrams, we expect that each additional
pair of rails will introduce a factor $1/\lambda^2$ from the imaginary part of
the HTL self energy. Since each additional chain vertex introduces a factor
$\lambda$, it appears naively that chain diagrams contain a factor
$(1/\lambda)^n$ where $n$ is the number of chain links. In fact this
conclusion is invalid. A pair of rails in a chain link does not contribute a
factor $1/\lambda^2$ because of the fact that the discontinuity of a chain link
vanishes in the limit of zero external four momentum, and the real part of a
bubblechain link does not contain pinching pole contributions. Consequently, chain
diagrams are suppressed by a factor $\lambda^n$ where $n$ is the number of
links. The four-point vertex is introduced only as a means to illustrate the
self-consistent cancellation, as discussed in section [VII].

\subsubsection{Integral equation for viscosity}

We need to express the viscosity in terms of an integral equation of the form
shown in Fig. [16]. For the first term we obtain,
\bea
\lambda^{(1)}_{ab} = -i\lambda \int dK\,dP\, I_{lm}(k,k) I_{lm}(k,k)
D_{ac}(P+Q) D_{ca}(P) \tau_3^c D_{cb}(K+Q) D_{bc}(K)
\eea
Performing the summations and taking the combination $\lambda^{(1)}_{21} -
\lambda^{(1)}_{12} = (\sigma - \sigma')^{(1)}$ we obtain,
\bea
&&(\sigma-\sigma')^{(1)} \label{intM1st}\\
&& = -i\frac{\lambda}{4} \int dK\,dP\, I_{lm}(k,k)
I_{lm}(k,k)[(N_p-N_{p+q})(N_k-N_{k+q})(r_pa_{p+q}r_ka_{k+q} + h.c.)] \nonumber
\eea
For the second diagram in Fig. [16] we obtain,
\bea
\lambda^{(2)}_{ab} =&& \int dK\, dP\, I_{lm}(p,p) M(P+Q,-K-Q,-P,K)_{eced}
\nonumber \\
&&D_{ea}(P) D_{ae}(P+Q) D_{bd}(K) D_{cb}(K+Q)I_{lm}(k,k) \nonumber
\eea
Taking the appropriate combination we get,
\bea
&&(\sigma-\sigma')^{(2)} = \frac{1}{8}\int dK\,dP\, I^{lm}(p,p)\nonumber \\
&&~~\left\{ \right.(N_{k+q} - N_k)r_{k+q} a_k [ M_B(r_p a_{p+q} + a_p r_{p+q} +
f_p f_{p+q}) + (M_{R1} + M_{R3})(a_p f_{p+q} + f_p r_{p+q})]\nonumber \\
&&~~+(N_{p+q}-N_p)r_p a_{p+q} (M_B(f_k f_{k+q} +a_k r_{k+q}) + (M_D + M_T)r_k
a_{k+q} + \nonumber\\
&&~~~~(M_{R4} + M_\beta)r_k f_{k+q} + (M_{R2} + M_\delta) f_k a_{k+q}
\left.\right\}
 I_{lm}(k,k)
\eea
This result can be written in terms of a three-point function.
Using~(\ref{KMSver}),~(\ref{decompM}) and~(\ref{KMSM}) we find,
\bea
&&M_T = M_D := iZ_F;~~~M_B:=0;~~~M_{R1} = M_{R3}:=iZ_R;\nonumber \\
&&~~~M_{R2} = M_\delta := iZ_{Ri};~~~M_{R4} = M_\beta :=iZ_{Ro} \nonumber
\eea
where the momentum arguments for the $M$'s and the $Z$'s are
$M_x(P+Q,-K-Q,-P,K)$ and $Z_x(-K-Q,Q,K)$ and the vertices $Z$ obey the same KMS
conditions (\ref{KMSver}) as the vertices $\Gamma_{lm}$.
Rewriting the result in terms of the three-point function $Z$ we find that the
viscosity can be expressed in terms of  one retarded three-point function:
\bea
(\sigma-\sigma')^{(2)} =&& \frac{i}{4} \int dK\,dP\,I^{lm}(p,p) (N_k-N_{k+q})
(N_p-N_{p+q})\label{intM2nd}\\
&&[a_k r_{k+q} a_p r_{p+q} Z_R(-K-Q,Q,K) + h.c.]I_{lm}(k,k)\nonumber
\eea
Combining~({\ref{intM1st}) and~(\ref{intM2nd}) we obtain,
\bea
\sigma-\sigma' =&& \frac{i}{4} \int dK\,dP\,I^{lm}(p,p) (N_k-N_{k+q})
(N_p-N_{p+q})\label{intM222}\\
&&(a_k r_{k+q} a_p r_{p+q} [-\lambda + Z_R(K,Q,-K-Q)] +
h.c.)I_{lm}(p,p)\nonumber
\eea

\subsubsection{The splitting relation for the four-point function}

We consider the one loop diagrams shown in Fig. [17a,b]. The vertex shown in
Fig. [17a] is given by,
\bea
&&M_{bcba}(P+Q,-K-Q,-P,K) \nonumber \\
&&~~~~= i\frac{\lambda^3}{2} \int dS\,dR\,D_{bc}(R+K+Q) D_{ab}(R+K) D_{ac}(S)
D_{ca}(R+S) \tau_a \tau_b \tau_c \nonumber
\eea
Using the form of the propagators given in~(\ref{decompD1}) and extracting the
vertex $M_{R1} = iZ_R$ we obtain,
\bea
Z_R(K,Q,-K-Q) =&& \frac{\lambda^3}{4} \int dS\,dR\,\left[ r_{r+k+q}
a_{r+k}\delta + a_{r+k} f_{r+k+q} \phi_1 + f_{r+k} r_{k+r+q} \phi_2 \right]
\eea
where $\delta,~~\phi_1,~~\phi_2$ are functions of the form $f(s,r+s)$.
Using~(\ref{splitprop}) to split pairs of propagators of the form $D_{r+k}
D_{r+k+q}$  we obtain,
\bea
Z_R = \frac{\lambda^3}{4}\frac{1}{Q^2+2Q\cdot K} \int dS\,dR [a_{r+k} \delta +
f_{r+k} \phi_2 - r_{k+r+q} \delta - f_{k+r+q} \phi_1 ]\nonumber
\eea
where we are looking at the region of phase space where $2Q\cdot R \ll Q^2 +
2Q\cdot K$.
We compare this expression with the results for the self energy shown in Fig.
[17b]. We find that,
\bea
Z_R = -\frac{3\lambda}{Q^2 + 2Q\cdot K}(\Pi_A(K) - \Pi_R(K+Q))
\label{Msplit}
\eea
The lowest order contribution to the viscosity is shown in Fig. [17c]. Notice
that the one loop graphs shown in Fig. [18] and the two loop graphs shown in
Fig. [19] do not contribute to the expression~(\ref{Msplit}) since the tadpole
self energies are momentum independent.
At higher loop order, the diagrams become increasingly complicated. Working at zero temperature for simplicity, the same relation is
satisfied by the diagrams in Figs. [20-24].

\subsubsection{An Integral equation for $\Pi$}

In order to perform a resummation of a series of contributions to the
viscosity, we need to obtain $\Pi$ as the solution of an integral equation
which resums contributions to the self energy. As in section [VI-A-2], we
consider the integral equation which resums sunset diagrams (see Fig. [13]).
In this case, none of the lines contain crosses, because we do not need to work
with a self energy that contains factors $I_{lm}(p,p)$. As explained in detail
below, this simplification is the motivation for introducing the four-point
vertex.
We find (compare~(\ref{pi-int})),
\bea
&&\Pi(K)_R = \Pi_0(K)_{R} -\frac{\lambda}{8}\int dR\,dP\,\left( \delta
\Pi(K+P)_R r_{k+p} r_{k+p}\right. \nonumber
\\
&&~~+\left. \phi_2[\Pi(K+P)_R f_{k+p} r_{k+p} + \Pi(K+P)_A a_{k+p} f_{k+p} +
\Pi(K+P)_F a_{k+p} r_{k+p}]\right) \nonumber \\
&&\Pi(K)_A =  \Pi_0(K)_{A} -\frac{\lambda}{8}\int dR\,dP\,\left( \delta
\Pi(K+P)_A a_{k+p} a_{k+p} \right.\label{pi-intM} \\
&& ~~+ \left.\phi_1[\Pi(K+P)_R f_{k+p} r_{k+p} + \Pi(K+P)_A a_{k+p} f_{k+p} +
\Pi(K+P)_F a_{k+p} r_{k+p}]\right) \nonumber
\eea

\subsubsection{Contributions to the viscosity}

As in section [VI-A-3], the last step is to ask what contributions to the
viscosity are obtained by substituting the solution of the integral
equation~(\ref{pi-intM}) into~(\ref{Msplit}) (which is in turn substituted
into~(\ref{intM222})). It is straightforward to show that the diagrams
produced are the same as for the case of the three-point vertex (as shown in
Fig. [13]), except for the fact that each diagram contains one bare chain link
insertion. One example is shown in Fig. [15]. From power counting arguments,
it is known that diagrams containing chain link contributions are suppressed.
We reiterate, that the four-point vertex is discussed only as a means to
illustrate the self consistent cancellation, as discussed in the next section.

\subsubsection{The Infrared Divergence}

As always, the result is infrared divergent unless the propagators are replaced
with HTL propagators. At lowest order, the diagrams that satisfy
\bea
\label{ZsplitHTL}
Z_R(K,Q,-K-Q) = -  \frac{1} {Q^2+ 2Q\cdot K -\Sigma_R(Q+K) + \Sigma_A(K)}
(\Pi_A(K)- \Pi_R(Q+K)) \eea
are the same as in Figs. [17a,b] with the propagators replaced by HTL
propagators. Performing the substitution we obtain, at first order, the
contribution to the viscosity shown in Fig. [25].

\section{Self Consistent Resummations}

It has been shown that the resummation technique introduced in Section
[VI] can be used to resum a larger class of diagrams than just the ladders. In
the case of the corrected three-point vertex, the ladder graphs are included in
the resummation, in addition to another large group of graphs that are shown to
contribute at the same order. In the case of the corrected four-point vertex,
we include the same large class of diagrams as before, expect that all
contributions to the viscosity will have a chain link piece to the diagram.
Now we want to consider a further increase in the set of diagrams to be
resummed. We follow \cite{meg1}. The basic idea is as follows. Until now, we
have considered various forms of vertex corrections, with the propagator lines
corrected by HTL self energies to obtain an infrared finite result. Now we
will consider diagrams with corrected propagators of the form,
\bea
\bar{D}(P) = \frac{1}{P^2 - \Pi(P)}\label{barprop}
\eea
where the self energies are not HTL self energies, but completely general
expressions. Our goal is to find a cancellation between the self energies in
expressions for the vertices ((\ref{g-p-lm}) and~(\ref{Msplit})), and the self
energies in the corrected propagators. In \cite{meg1}, in a calculation of the
self energy in $\phi^3$ theory, it has been shown that these cancellations occur
and are independent of the form of the self energy. The value of this approach
is that the vertex corrections and the propagator corrections are written in
the same form and dealt with simultaneously and, in this sense, they are
treated self-consistently.

We show below using the four-point function that it is possible to obtain a
partial cancellation of the type described above. We remind the reader that in
addition to the fact that the cancellation is not complete, there are other
problems with the four-point vertex approach. As explained in section [VI-B],
the decomposition of the four-point function in terms of self energies cannot
be accomplished in general. We must restrict to four-point functions in which
two of the fields are at the same space point, and the corresponding two
Keldysh indices are equal. The consequence of this restriction is that we only
obtain contributions to the viscosity in which there is one chain link piece.
It seems likely that these two problems are related. We use the four-point
vertex only as a means to illustrate the cancellation in the scalar theory.

We begin with the integral equation shown in Fig. [26]. We have
(compare~(\ref{intM222})),
\bea
\sigma-\sigma' = \frac{i}{4} \int &&dK\,dP\,
(N_p-N_{p+q})(N_k-N_{k+q})I_{lm}(p,p) \nonumber\\
&&\{ a_{p+q} r_p \bar{a}_{k+q} \bar{r}_k (-\lambda +Z_R(-K-Q,Q,K)) + h.c.\}
I_{lm}(k,k)\label{Mintbar}
\eea
The propagators with bars are full propagators (\ref{barprop}), and the two
bare propagators will give the bare chain link piece to each viscosity diagram,
as discussed previously. Substituting in~(\ref{ZsplitHTL}) we obtain,
\bea
&&\sigma - \sigma' =  -\frac{i\lambda}{4} \int dK\,dP\,(N_k - N_{k+q} ) (N_p -
N_{p+q}) I_{lm}(p,p) I_{lm}(k,k) \nonumber \\
&&~~\left[ \left(1+\frac{3}{Q^2 + 2Q\cdot K - \Sigma_R(Q+K) +
\Sigma_A(K)}(\Pi_A(K) - \Pi_R(Q+K))\right) a_p r_{p+q} \bar{a}_k \bar{r}_{k+q}
+ h.c. \right] \nonumber \\
&&= -i \frac{\lambda}{4} \int dK\,dP\,\left(\frac{(N_k - N_{k+q})(N_p -
N_{p+q})}{Q^2 + 2Q\cdot K - \Sigma_R(Q+K) + \Sigma_A(K)}~~I_{lm}(p,p)
I_{lm}(k,k)\right.\nonumber \\
&&~~\left.[Q^2 + 2Q\cdot K - 3\Pi_R(Q+K) + 3\Pi_A(K) - \Sigma_R(Q+K) +
\Sigma_A(K) ] a_p r_{p+q} \bar{a}_k \bar{r}_{k+q} + h.c. \right)\nonumber
\eea
Using $\bar{D}(P)^{-1} = P^2 - \Pi(P)$ we obtain,
\bea
&&\sigma - \sigma' = -i \frac{\lambda}{4} \int dK\,dP\,\frac{(N_k - N_{k+q})
(N_p - N_{p+q})}{Q^2 + 2Q\cdot K - \Sigma_R(Q+K) + \Sigma_A(K)} a_p
r_{p+q}I_{lm}(p,p) I_{lm}(k,k) \nonumber\\
&&~~~~ \left(\bar{a}_k - \bar{r}_{k+q} + [2\Pi_A(K) - 2\Pi_R(Q+K) + \Sigma_A(K)
- \Sigma_R(Q+K)]\bar{a}_k \bar{r}_{k+q} + h.c.\right)\label{98}
\eea
When $\Pi$ is obtained as the solution of the integral equation
(\ref{pi-intM}), Eqn. (\ref{98}) represents the resummation of a huge number of
contributions to the viscosity. Unfortunately, the cancellation between the
vertex and propagator factors is only partial. We expect that the cancellation
will be complete in a gauge theory with a three-point interaction like scalar
QED. The motivation for this argument is discussed below.

When calculating viscosity from the corrected three-point function, the
presence of the composite operator~(\ref{ppi}) leads to difficulties. As shown
in section [VI-A-1], this composite operator leads to a three-point vertex
which cannot be decomposed in terms of a standard self energy, but rather the
modified self energy which we have written as $\Pi_{lm}$. Even after using
translation invariance to write $\Pi_{lm} = I_{lm}\bar{\Pi}$, there is in
general no cancellation between the factor $\bar{\Pi}$ from the vertex and the
factor $\Pi$ from the full propagator. Note however that the numerical factor
seems to be correct. Consider substituting (\ref{g-p2}) into (compare
(\ref{int2})),
\bea
&&\sigma-\sigma' \label{int22} \\
&&=  \int dK\, \left\{ (I_{lm}(k,k) + \frac{1}{2}\Gamma^{lm}_R(K,Q,-K-Q))\bar{a}_k
\bar{r}_{k+q} + h.c.\right\} (N_{k+q} - N_k) I_{lm}(k,k) \nonumber
\eea
and replacing $\Pi_{lm}$ by $I_{lm} \bar{\Pi}$. If we could write
$\Pi=\bar{\Pi}$ then the cancellation between the factors of $\Pi$ in the
propagators and the vertex would be complete: there would be no factors of
$\Pi$ in the numerator of the resulting expression.

We expect that the cancellation that we have described in this section does
occur in a gauge theory like scalar QED as a result of the additional
constraints imposed by gauge invariance. Note that equations of the
form~(\ref{g-p}) look very much like Ward identities. It has been shown
\cite{meg1} that the corresponding equation for scalar QED is
\bea
\Gamma_R^\mu = -ie\frac{(Q+2K)^\mu}{Q^2+2Q\cdot K} (\Sigma_A(K) -
\Sigma_R(K+Q))
\eea
and corresponds to the statement that the part of the vertex that is large in
the infrared limit must satisfy the usual Ward identity with the polarization
tensor. In a future publication, we will look at this type of cancellation in
viscosity in scalar QED.

\section{Conclusions}

We have introduced several different techniques for performing non-perturbative
resummations. We used the closed time path to reformulate the traditional resummation of ladder
graphs and obtained the same integral equation as that obtained previously in
the imaginary time formalism. We have developed a technique which uses a pair
of integral equations that gives resummations of both ladder and non-ladder
graphs. We have argued that these non-ladder graphs contribute at leading
order. Finally, we have shown how this resummation technique can be
generalized to treat vertex and propagator corrections self-consistently in a
way that gives rise to some cancellation
 between vertex and propagator corrections.
We might expect that, in a gauge theory, the constraints imposed by the Ward
identities will facilitate this cancellation.

The various resummation schemes that we introduce in this paper become
increasingly difficult to understand in terms of diagrams. It seems likely
that a more physical understanding of these resummations can be obtained by
interpreting the resulting integral equations in terms of an effective kinetic
theory description. The equivalence of the integral equation corresponding to
the ladder resummation and the Boltzmann equation was first discussed in
\cite{jeon1}. It should be possible to obtain a similar interpretation of the
integral equations that correspond to the resummation schemes that we have
introduced.
%%%%%%%%%%%%%%%%%%%%%%%%%%%%%%%%%%%%%%%%%%%%%%%%%%%%%%%%%%%%
\acknowledgments
%%%%%%%%%%%%%%%%%%%%%%%%%%%%%%%%%%%%%%%%%%%%%%%%%%%%%%%%%%%%

 This work was supported by the Natural Sciences and Engineering 
and Research Council of Canada (NSERC), and the National Natural Science Foundation of China (NSFC).

\newpage

\Large

\centerline{\bf Appendix A}

\normalsize

\vspace*{1cm}

We start from the integral equation involving the three-point function
$\Gamma_{lm}$ (\ref{int2}). We work at zero temperature and consider only
the term that corresponds to corrections to the one loop graph. Note that, at zero temperature, it is not possible to regulate the pinch singularity using the imaginary part of the HTL self energy.   We use the zero temperature case for illustrative purposes only, to simplify the discussion at higher loop orders.
At zero
temperature we have,
\bea
\sigma = 2\int dK\, \Gamma_{lm}(K,Q,-K-Q) iD(K) iD(K+Q) I_{lm}(k,k)
\label{vis0T}
\eea
We use the zero temperature version of~(\ref{g-p-lm}),
\bea
\Gamma_{lm}(K,Q,-K-Q) \simeq \frac{2}{Q^2+2Q\cdot K}(\Pi_{lm}(K) -
\Pi_{lm}(Q+K))
\label{gamma0T}
\eea
 We make the definitions,
\bea
&& a= K-P_1-P_2 \nonumber \\
&& a' =  K+Q-P_1-P_2 \nonumber \\
&& b = K-P_1-P_2+P_3+P_4 \nonumber \\
&& b' = K+Q-P_1-P_2+P_3+P_4 \nonumber
\eea
and we write $D(a) = D_a$, etc.
We make throughout the approximation
\bea
I_{lm}(a,a) \simeq I_{lm}(a',a') \simeq I_{lm}(b,b) \simeq I_{lm}(b',b')
\nonumber
\eea
since there is no difficulty with taking quantities in the numerator to zero.

We look
at the self energy shown in Fig. [27]. We
obtain,
\bea
\Pi_{lm}(K) = -\frac{\lambda^4}{4}\int dP_1\,dP_2\,dP_3\,dP_4\,  D(P_1) D(P_2)
D(P_3) D(P_4)\, D_a^2 \,D_b \,I_{lm}(b,b) \nonumber
\eea
Substituting into~(\ref{vis0T}) and (\ref{gamma0T}) we have,
\bea
\sigma = \frac{\lambda^4}{2} &&\int dK\, dP_1\,dP_2\,dP_3\,dP_4\,
\frac{1}{Q^2+2Q\cdot K} \nonumber\\
&& D(P_1) D(P_2) D(P_3) D(P_4) I_{lm}(b,b) (D_a^2D_b-D_a'^2 D_b')   D(K) D(K+Q)
I_{lm}(k,k) \nonumber
\eea
We can rewrite this equation by using the following identity
\bea
D_a^2D_b-D_a'^2 D_b' = D_a D_b(D_a-D_a') + D_a' D_b'(D_a-D_a') + D_a
D_a'(D_b-D_b') \nonumber
\eea
The next step is to use the splitting relation for the propagators to write,
\bea
\frac{1}{Q^2 + 2Q\cdot K} (D_a-D_a') = D_a D_a';~~~~~~{\rm etc.} \nonumber
\eea
We obtain,
\bea
\sigma =  \frac{\lambda^4}{2} &&\int dK\, dP_1\,dP_2\,dP_3\,dP_4\,  D(P_1)
D(P_2) D(P_3) D(P_4) D(K) D(K+Q) \nonumber \\
&& [I_{lm}(a',a) D_a^2 D_a' D_b + I_{lm}(a',a) D_a D_a'^2 D_b' + I_{lm}(b',b)
D_a D_a' D_b D_b']   I_{lm}(k,k+q) \nonumber
\eea
which corresponds to the diagrams in Fig. [13b].
This procedure can also be extended to higher orders. It can be shown that the
polarization tensor in Fig [28] corresponds to the contributions to the
viscosity shown in Fig. [29].

\newpage
\listoffigures
%\newpage
 \begin{figure}
\epsfxsize=4cm
\centerline{\epsfbox{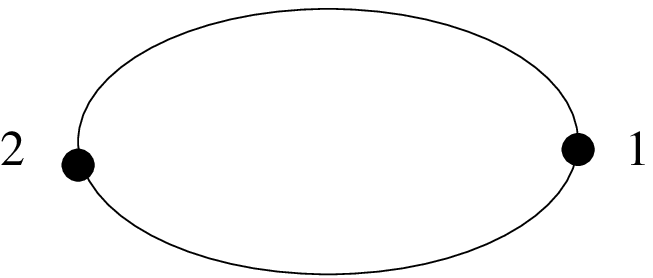}}
\vskip 0.4cm
 \caption{One-loop skeleton diagram for shear viscosity in $\phi^4$ theory.}
 \label{F1}
 \end{figure}

 \begin{figure}
\epsfxsize=6cm
\centerline{\epsfbox{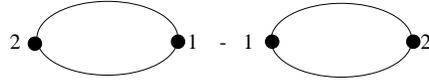}}
\vskip 0.4cm
\caption{Symmetric combination of one-loop diagrams for the shear viscosity 
  in $\phi^4$ theory.}
 %\label{F2}
 \end{figure}

 \begin{figure}
\epsfxsize=10cm
\centerline{\epsfbox{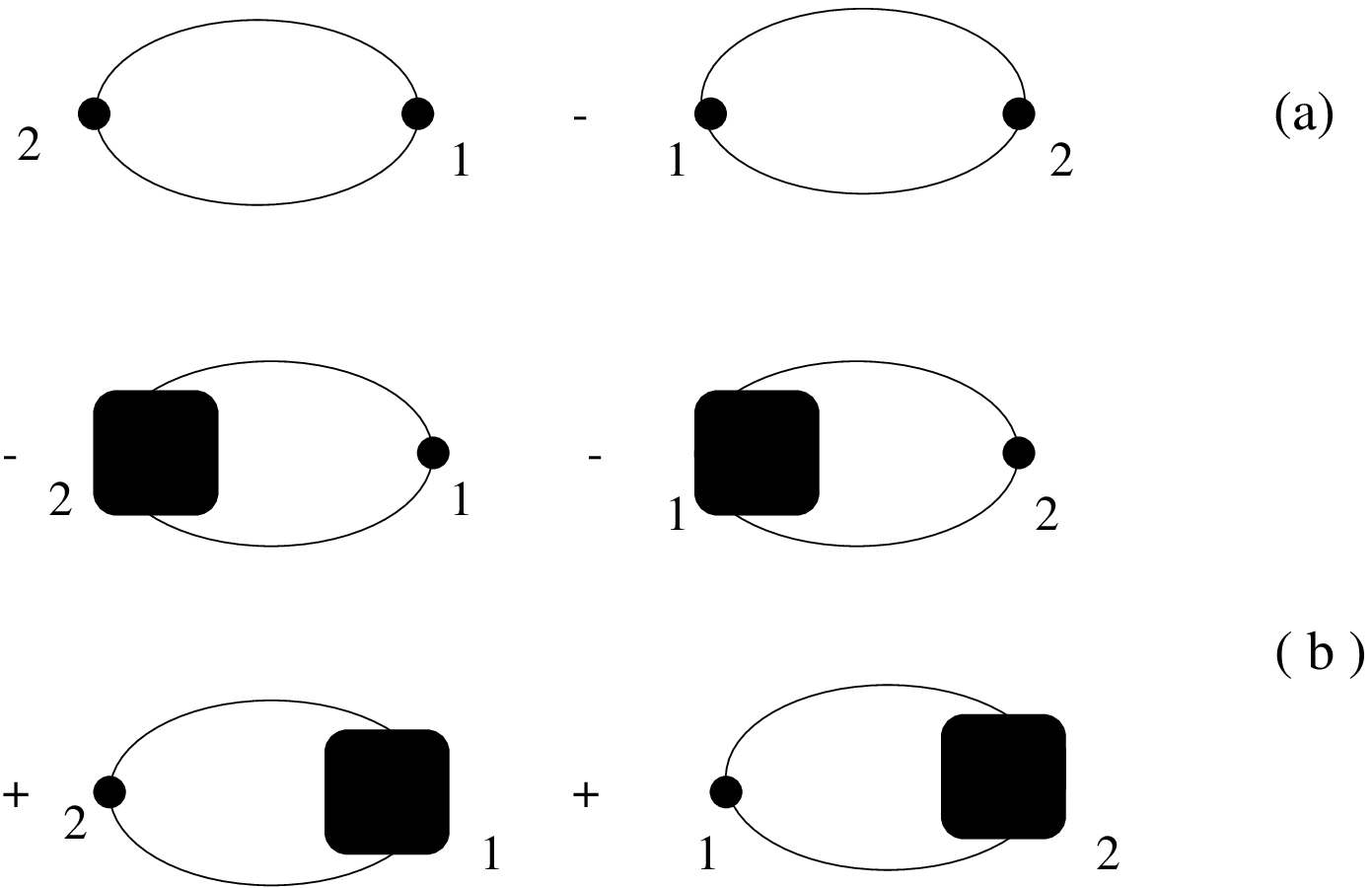}}
\vskip 0.4cm
\caption{Symmetric diagrams for the  shear viscosity in terms of a corrected 
three-point function.  The squre blob respresents the corrected three-point 
vertex.}
\label{F3}
 \end{figure}

 \begin{figure}
\epsfxsize=6cm
\centerline{\epsfbox{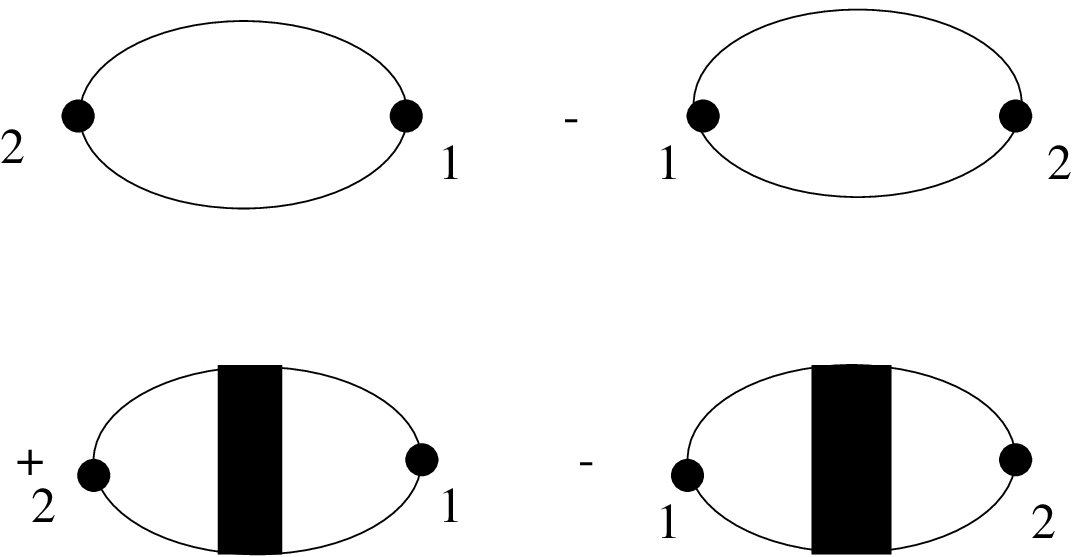}}
\vskip 0.4cm
\caption{Symmetric diagrams for the shear viscosity 
in terms of a corrected four-point function. The rectangle respresents 
the corrected four-point 
vertex.}
\label{F4}
 \end{figure}
\newpage

 \begin{figure}
\epsfxsize=4cm
\centerline{\epsfbox{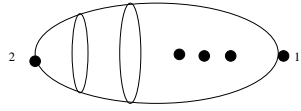}}
\vskip 0.4cm
 \caption{The infinite series of  planar-ladder diagrams in $\lambda\phi^4$
theory.}
 \label{F5}
 \end{figure}
%\newpage

\begin{figure}
\epsfxsize=6cm
\centerline{\epsfbox{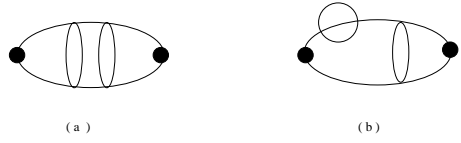}}
%\vskip 0.4cm
 \caption{ Five-loop ladder- (a) and non-ladder-graph (b) for the shear viscosity.}
 \label{F6}
 \end{figure}

\begin{figure}
\epsfxsize=8cm
\centerline{\epsfbox{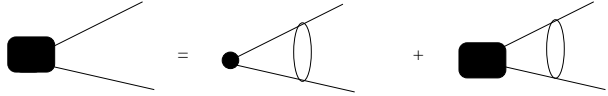}}
\vskip 0.4cm
 \caption{Schwinger-Dyson equation for the corrected three-point vertex 
which includes  ladder graphs  in $\phi^4$ theory.}
 \label{F7}
 \end{figure}

 \begin{figure}
\epsfxsize=8cm
\centerline{\epsfbox{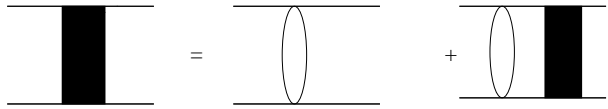}}
\vskip 0.4cm
 \caption{Schwinger-Dyson equation for corrected four-point vertex which  
includes  ladder graphs  in $\phi^4$ theory.}
 \label{F8}
 \end{figure}

 \begin{figure}
\epsfxsize=6cm
\centerline{\epsfbox{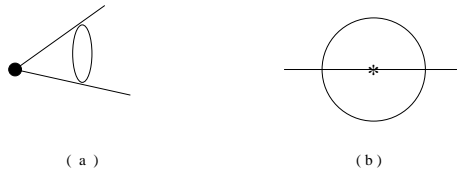}}
\vskip 0.4cm
 \caption{(a) The three-point vertex including one ladder insertion;  
(b)the sunset self-energy that corresponds  to (a); the asterix represents a factor $I_{lm}$.}
\label{F9}
 \end{figure}
\newpage

 \begin{figure}
\epsfxsize=8cm
\centerline{\epsfbox{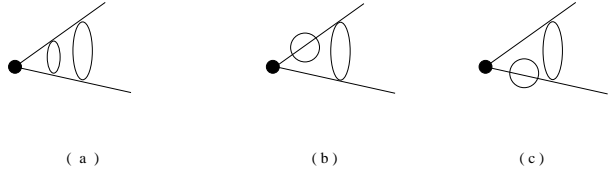}}
\vskip 0.4cm
 \caption{(a) The three-point vertex with  two  ladders; (b) (c)   
non-ladder diagrams which contribute at the same order. }
\label{F10}
 \end{figure}

\begin{figure}
\epsfxsize=4cm
\centerline{\epsfbox{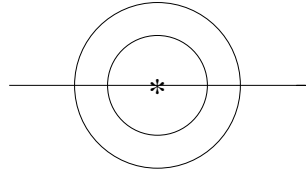}}
\vskip 0.4cm
 \caption{The second order sunset graph that corresponds to  Fig. [10]. }
\label{F11}
 \end{figure}

\begin{figure}
\epsfxsize=8cm
\centerline{\epsfbox{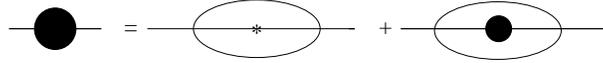}}
\vskip 0.4cm
 \caption{The integral equation that resums sunset contributions to the self-energy.  }
\label{F12}
 \end{figure}

\begin{figure}
\epsfxsize=10cm
\centerline{\epsfbox{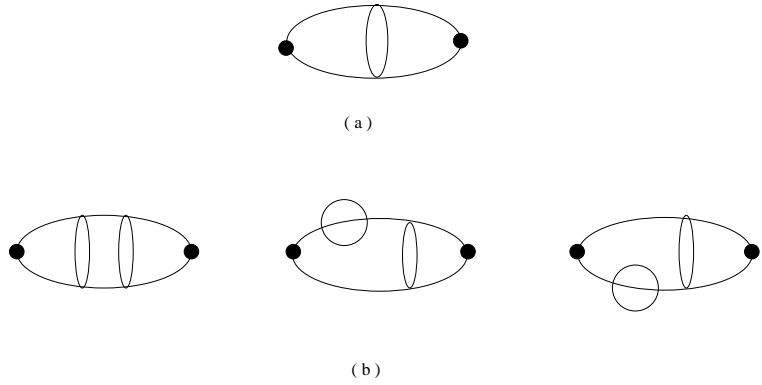}}
\vskip 0.4cm
 \caption{(a) The first order  and (b) the second order contributions to the shear viscosity.}
\label{F13}
 \end{figure}

\begin{figure}
\epsfxsize=4cm
\centerline{\epsfbox{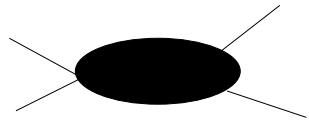}}
\vskip 0.4cm
 \caption{Corrected four-point vertex with two joined legs.}
\label{F14}
 \end{figure}
\newpage
\begin{figure}
\epsfxsize=4cm
\centerline{\epsfbox{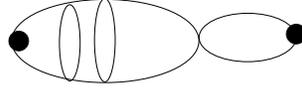}}
\vskip 0.4cm
 \caption{Typical ladder diagram  for the shear viscosity with one chain.}
\label{F15}
 \end{figure}

\begin{figure}
\epsfxsize=6cm
\centerline{\epsfbox{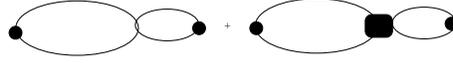}}
\vskip 0.4cm
 \caption{ Contribution  one chain  to the shear viscosity. Ihe black square represents the corrected four-point vertex.}
\label{F16}
 \end{figure}

\begin{figure}
\epsfxsize=6cm
\centerline{\epsfbox{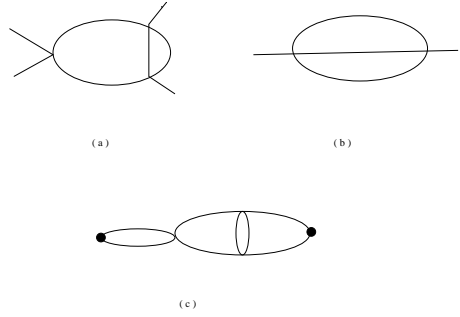}}
\vskip 0.4cm
 \caption{(a) The four-point vertex  and (b)its corresponding   sunset diagram; (c)The
one-chain and one-ladder diagram for the viscosity. }
\label{F17}
 \end{figure}

\begin{figure}
\epsfxsize=6cm
\centerline{\epsfbox{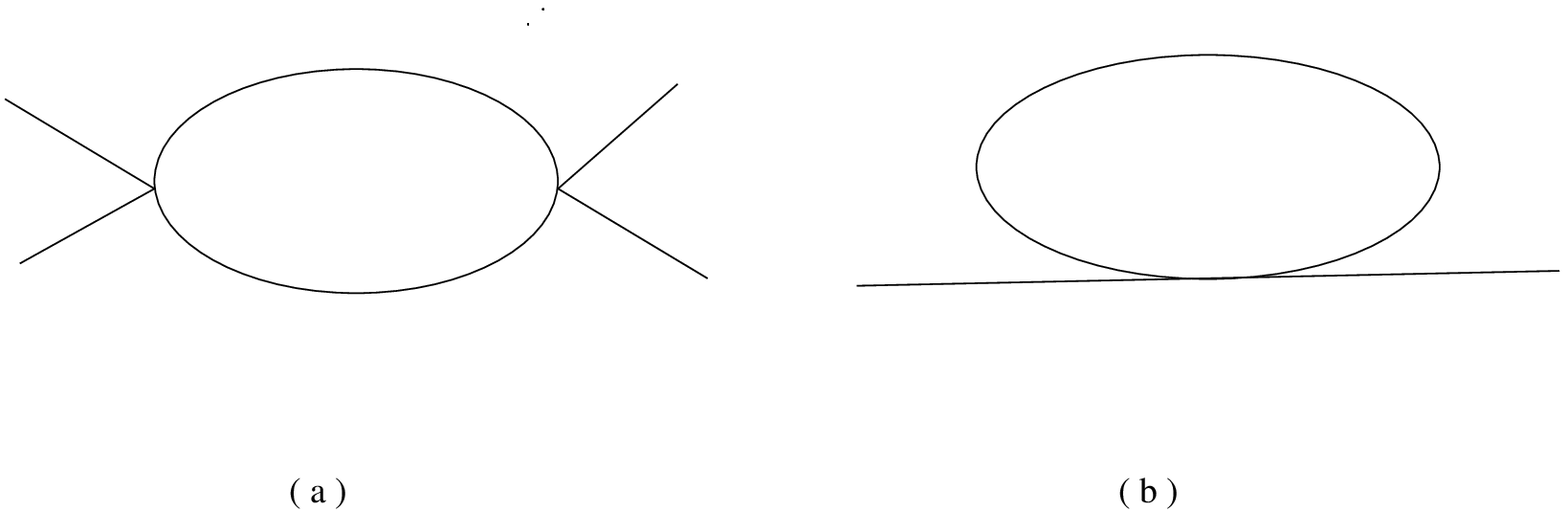}}
\vskip 0.4cm
 \caption{(a) The one-loop four-point vertex  and (b) the corresponding 
self-energy. }
\label{F18}
 \end{figure}

\begin{figure}
\epsfxsize=6cm
\centerline{\epsfbox{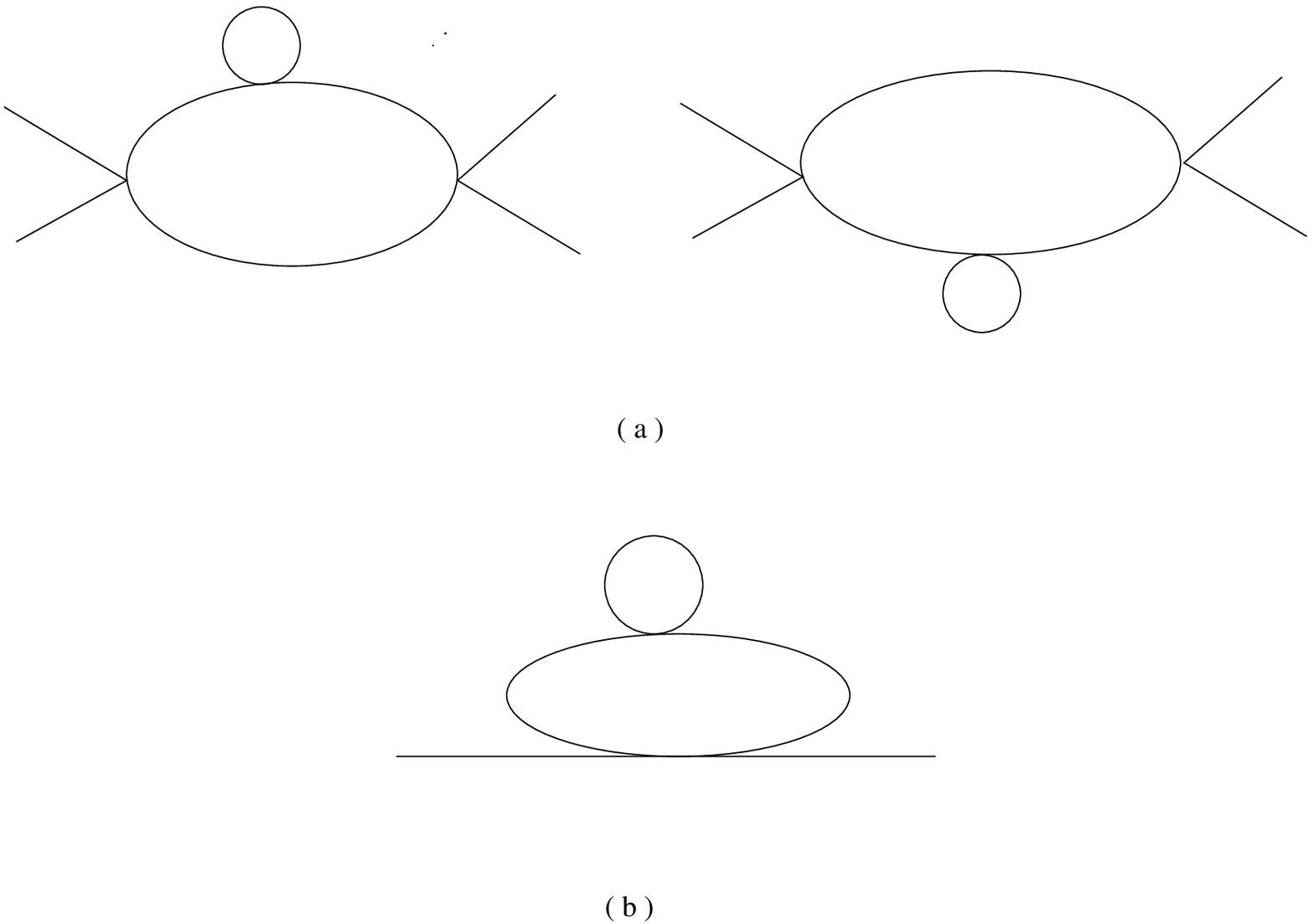}}
\vskip 0.4cm
 \caption{ (a) Some two-loop four-point vertices and (b) their corresponding
  self-energy.  }
\label{F19}
 \end{figure}

\begin{figure}
\epsfxsize=8cm
\centerline{\epsfbox{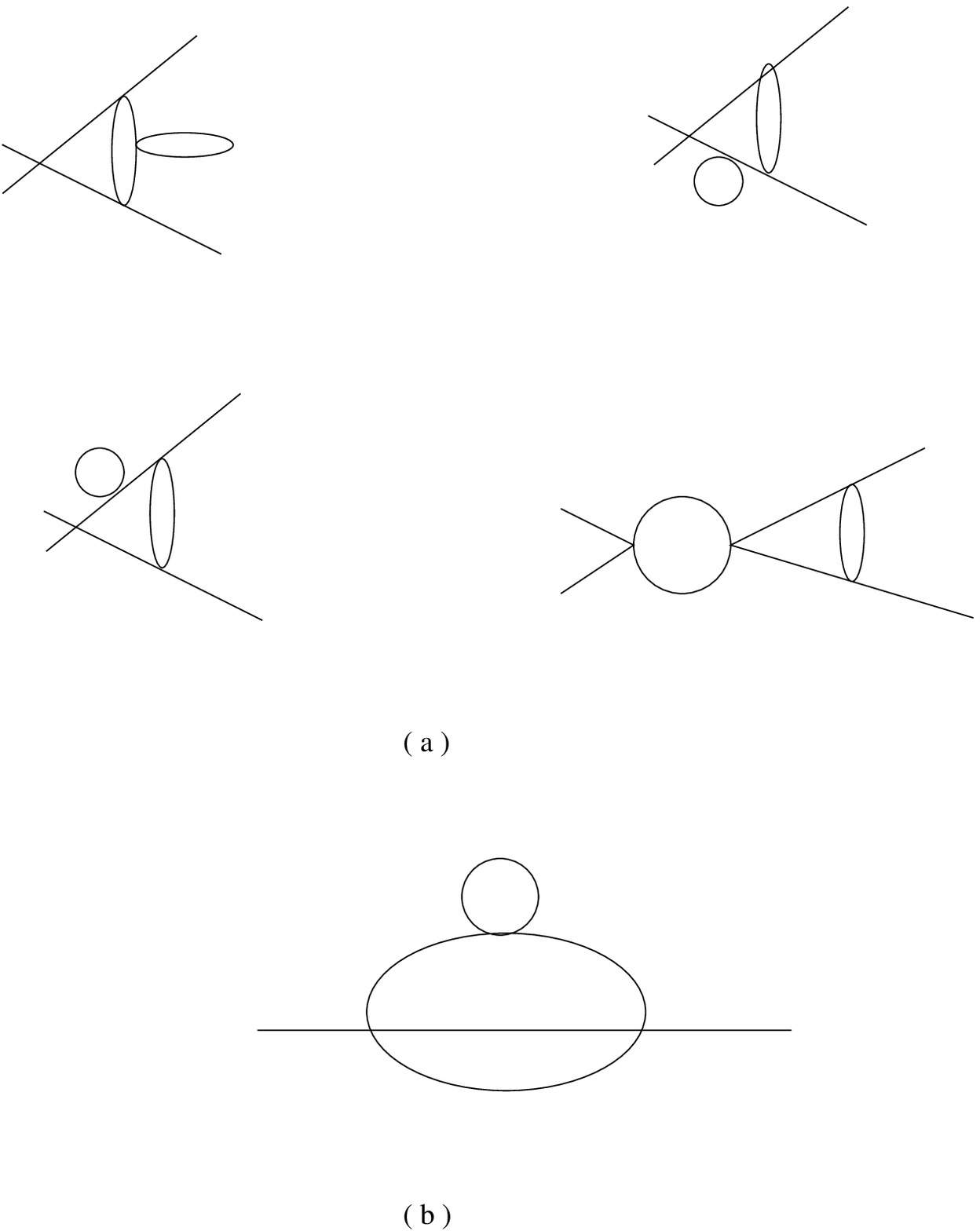}}
\vskip 0.4cm
 \caption{ (a) Three-loop four-point vertices with ladders   and (b) the
 corresponding sunset diagram. }
\label{F20}
 \end{figure}

\begin{figure}
\epsfxsize=6cm
\centerline{\epsfbox{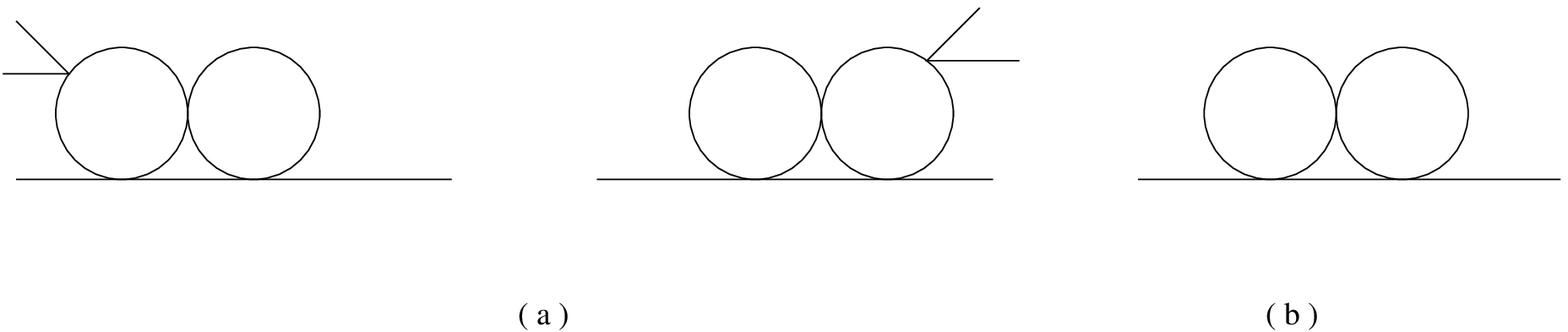}}
\vskip 0.4cm
 \caption{ (a)  Three-loop four-point vertices   and (b) the
 corresponding  self-energy. }
\label{F21}
 \end{figure}

\begin{figure}
\epsfxsize=6cm
\centerline{\epsfbox{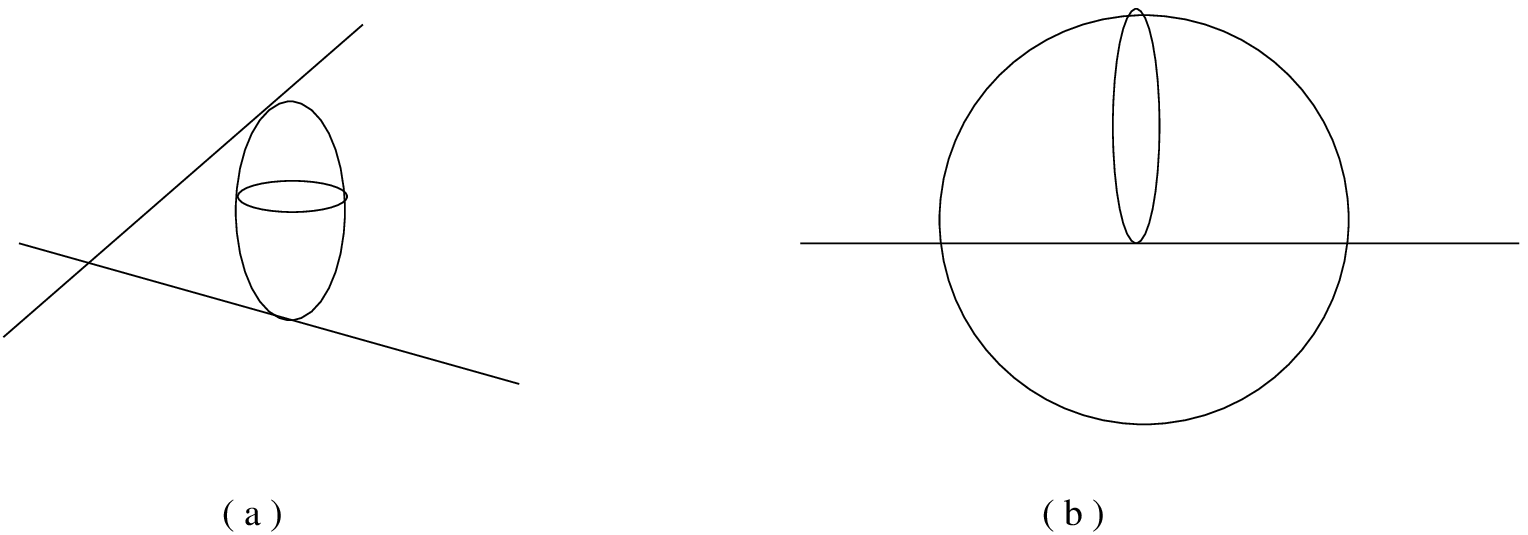}}
\vskip 0.4cm
 \caption{(a) Four-loop four-point vertex  and (b) the
 corresponding self-energy. }
\label{F22}
 \end{figure}

\begin{figure}
\epsfxsize=8cm
\centerline{\epsfbox{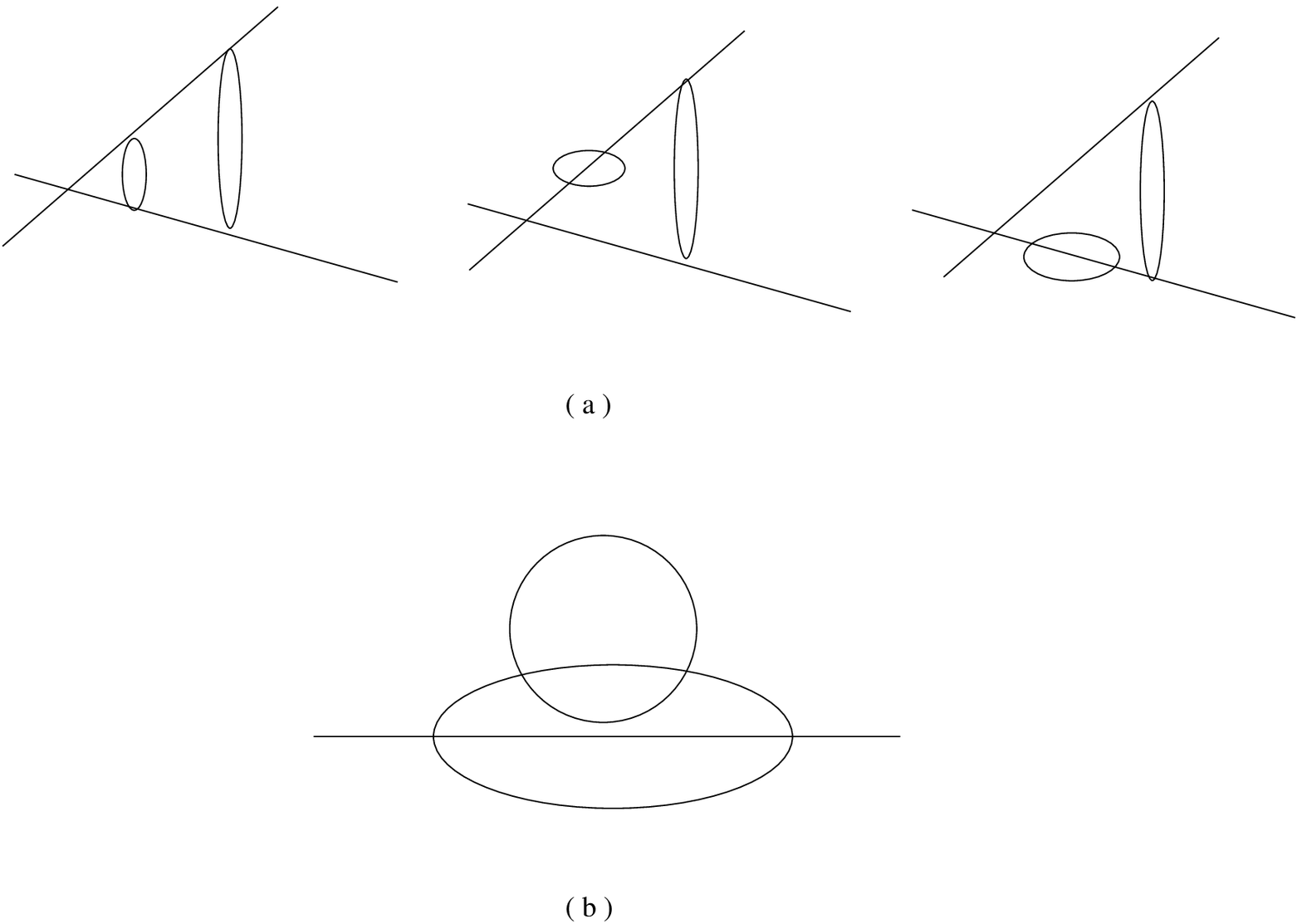}}
\vskip 0.4cm
 \caption{ (a) Four-loop four-point vertices and (b) the
 corresponding self-energy. }
\label{F23}
 \end{figure}

\begin{figure}
\epsfxsize=10cm
\centerline{\epsfbox{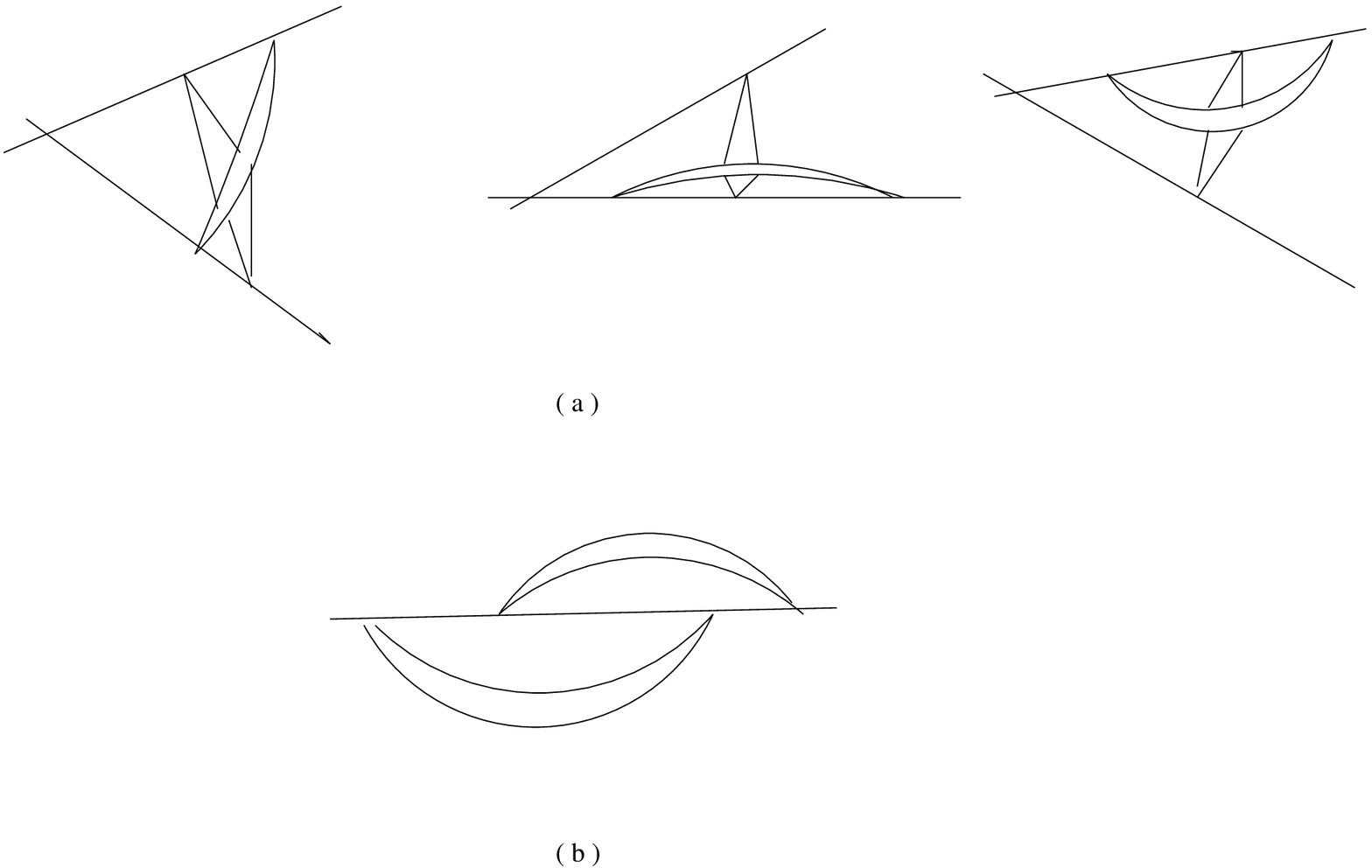}}
\vskip 0.4cm
 \caption{ Four-loop four-point vertices  and (b) the corresponding
  self-energy. }
\label{F24}
 \end{figure}

\newpage
\begin{figure}
\epsfxsize=6cm
\centerline{\epsfbox{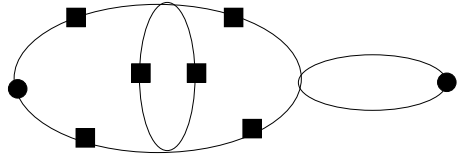}}
\vskip  0.4 cm
 \caption{ The shear viscosity with one-chain link and one ladder. The box 
indicates an effective propagator. }
\label{F25}
 \end{figure}

\begin{figure}
\epsfxsize=6cm
\centerline{\epsfbox{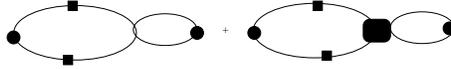}}
\vskip - 1cm
 \caption{ The shear viscosity with corrected 4-point vertex for diagrams with one chain link and  effective propagators. }
\label{F26}
 \end{figure}

%\newpage

\begin{figure}
\epsfxsize=4cm
\centerline{\epsfbox{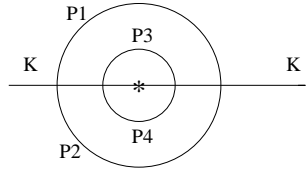}}
\vskip 0.4cm
 \caption{ The second order sunset diagram. }
\label{F27}
 \end{figure}

%\begin{figure}
%\epsfxsize=4cm
%\centerline{\epsfbox{fmm28.eps}}
%\vskip 0.4cm
% \caption{ A second order sunset diagram. }
%\label{F28}
% \end{figure}

%\newpage

\begin{figure}
\epsfxsize=4cm
\centerline{\epsfbox{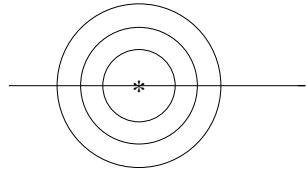}}
\vskip 0.4cm
 \caption{ The third order sunset diagram. }
\label{F28}
 \end{figure}

\begin{figure}
\epsfxsize=8cm
\centerline{\epsfbox{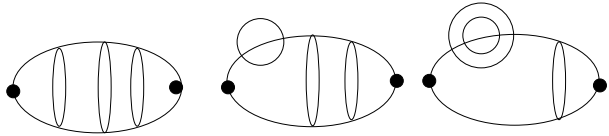}}
\vskip 0.4cm
 \caption{ Some of the ladder and non-ladder diagrams corresponding to the third order sunset diagram.}
\label{F29}
 \end{figure}

\end{document}